\documentstyle[aps,prb,multicol,eqsecnum,epsf]{revtex}
\draft
\def\({\begin{equation}}
\def\){\end{equation}}
\begin{document}                
\title{Theory of Interaction Effects in N-S Junctions out of Equilibrium}
\author{B. N. Narozhny and I. L. Aleiner}
\address{Department of Physics and Astronomy,
SUNY Stony Brook, Stony Brook, NY 11794}
\author{B. L. Altshuler}
\address{Department of Physics, Princeton University, Princeton, NJ 08544}
\address{NEC Research Institute, 4 Independence Way, Princeton, NJ 08540}
\maketitle
\begin{abstract}
We consider a normal metal - superconductor (N-S) junction in the regime,
when electrons in the normal metal are driven out of equilibrium.  We show
that the non-equilibrium fluctuations of the electron density in the
N-layer cause the fluctuations of the phase of the order parameter in the
S-layer.  
As a
result, the density of states in the superconductor deviates
from the BCS form, most notably the density of states
in the gap becomes finite.
This effect can be viewed as a
result of the time reversal symmetry breaking due to the non-equilibrium,
and can be described in terms of a low energy collective mode of
the junction, which couples normal currents in N-layer and supercurrents. 
This mode  is
analogous to the Schmid-Sch\"{o}n mode.
To interpret their measurements of the tunneling current, 
Pothier {\em et. al} [Phys. Rev. Lett. {\bf 79}, 3490 (1997)]
had to assume that the energy relaxation rate in the normal metal
is surprisingly high.  The
broadening of the BCS singularity of the density of states in the S-layer
manifest itself similarly to the broadening of the distribution function.
Mechanism suggested  here can be a possible explanation of this 
experimental puzzle.  We also propose an independent 
experiment to test our explanation.
\end{abstract}

\begin{multicols}{2}
\narrowtext

\section{Introduction}

In  metals the inelastic scattering rate  $1/\tau_{in}$
at low enough energies is determined by electron-electron interactions.
In a clean Fermi liquid  $\epsilon \tau_e \gtrsim \hbar$ this
inelastic rate can be estimated as
$\hbar/\tau_{in} \simeq \epsilon^2/\epsilon_F$, 
where $\epsilon$ is the
energy of the quasiparticle, $\tau_e$ is the elastic scattering time,
and $\epsilon_F$ is the Fermi energy.
This familiar result reflects only the phase volume
of the final state for an inelastic process,
while the corresponding matrix element 
is an energy independent constant.
In the dirty limit $\epsilon \tau_e < \hbar$ 
the inelastic rate is significantly
enhanced as compared with the clean case 
due to long range diffusive correlations of single
electron wave-functions in the disordered system\cite{relaxation,altr},
see Refs.~\onlinecite{functions} for more detailed
discussion.

The inelastic scattering rate is not by itself an observable quantity. 
However, inelastic collisions of electrons 
profoundly affect the behavior of the
system and,  therefore, $1/\tau_{in}$ 
in many cases can be extracted from experimental data. 
For example, from magnetoresistance one can evaluate
quantitatively the dephasing time $\tau_{\varphi}$, 
which often coincides with $1/\tau_{in}$.
The dephasing time describes the
loss of phase coherence, as electrons move diffusively 
in the bulk of a metallic sample.
This loss of coherence cuts off 
otherwise divergent weak-localization correction. 
The dephasing time has been extensively studied, 
and we believe that the existing theory allows a
good understanding of the experimental data\cite{altr}.

Another effect of inelastic scattering is the energy relaxation
described by the time $\tau_\epsilon$.
This is the time it takes for a "hot" quasiparticle 
with energy $\epsilon$ much larger 
than temperature $T$ to thermalize with all the other electrons. 
Theoretically, it is given by\cite{relaxation}

\begin{equation}
\frac{\hbar}{\tau(\epsilon)} \simeq \frac{\epsilon}{G_{m}(L_\epsilon)};\quad
L_\epsilon = \sqrt{{\hbar D}\over{\epsilon}},
\label{tauep}
\end{equation}

\noindent
where $G_{m}(L) \propto L^{d-2} $ is the
dimensionless (in units of $e^2/2\pi \hbar$) 
conductance of a sample of size $L$,
$D$ is the diffusion constant and $d$ the dimensionality of the system.
This result follows from the Fermi Golden Rule, 
but now the phase volume is multiplied by the matrix element, 
which is no longer a constant. 
Not only this matrix element substantially depends on the energy
transferred, but it even
diverges at small energies due to the wave function correlation, 
see Refs.~\onlinecite{functions}.

To determine $\tau_\epsilon$ experimentally,
one has to apply an external perturbation to drive the system out of
equilibrium, and then to measure the distribution function of electrons.
Recently, an elegant and important experiment\cite{exp1} 
was performed to measure directly the 
electronic distribution function $f(\epsilon)$. 
An external voltage $U$ applied to a copper wire caused an
electric current $J$, thus driving the wire 
out of equilibrium.
In order to determine the non-equilibrium distribution function
the authors of Ref.~\onlinecite{exp1} fabricated
an additional electrode connected with the wire
by a tunneling contact.
The distribution function $f(\epsilon)$ was extracted from the 
measurements of the tunneling conductance $G_T(V)$ of this contact
as a function of the bias voltage $V$ 
using the procedure as follows. 
Assuming that the density of electronic states in the
wire is energy independent, one can present the 
tunneling conductance $G_T(V)$ as the convolution of $f(\epsilon)$  
with the tunneling density of
states in the additional electrode $\rho(\epsilon)$

\begin{equation}
G_T(V) \propto \int d\epsilon
{{\partial f(\epsilon - eV)}\over{\partial\epsilon}} \rho(\epsilon).
\label{Gt}
\end{equation}

\noindent
As follows from Eq.~(\ref{Gt}), the more pronounced is the energy 
dependence of the
density of states in the additional electrode $\rho(\epsilon)$,
the more precisely one can determine 
the distribution function $f(\epsilon)$ measuring $G_T(V)$. 
For this reason the authors of Ref.~\onlinecite{exp1} used a
superconducting electrode to take advantage
of the BCS singularity in the density of states $\rho(\epsilon)$.
The existence of this singularity in the equilibrium
was convincingly determined by independent measurements at $J=0$.
The data on the tunneling conductance $G_T(V)$ 
were fitted by Eq.~(\ref{Gt}) yielding the distribution function
$f(\epsilon)$ and, thus, the energy relaxation time.

This procedure produced quite unexpected results. 
First of all, the extracted relaxation time 
turned out to be two orders of
magnitude shorter than that of Eq.~(\ref{tauep}). 
Moreover,  no dependence 
of the relaxation time on the energy $\epsilon$ was observed.   
This would mean the failure of 
the theory lying behind the derivation of Eq.~(\ref{tauep}). 

However, the theory is based on regular expansion in
inverse powers of the dimensionless conductance $G_{m}$ 
(see Ref.~\onlinecite{functions} and references therein).  
The observation of a good
metallic conductance $G_{m} \gg 1$ in the experiment\cite{exp1}
justifies the applicability of this theory.

In this paper we attempt to explain the puzzling results 
of the experiment \cite{exp1} by lifting the main assumption 
in the interpretation of the data -- independence 
of the density of states in the superconductor 
of the electronic distribution in the normal metal.  
We  calculate the tunneling conductance 
between  the superconducting and metallic films  explicitly and
find that  interaction of the
tunneling electrons with non-equilibrium fluctuations 
of the  current in the
normal layer smears the BCS singularity 
in the density of states in the superconductor. 
This effect has nothing to do with energy relaxation.
Nevertheless, it effectively broadens the energy
dependence of the experimental $f(\epsilon)$, extracted from  
Eq.~(\ref{Gt})
with the use of the  equilibrium BCS density of states $\rho(\epsilon)$,
whereas the real distribution function remains sharp. 
As a result, the energy relaxation time appears much shorter,
than it really is.
We found that this effect is not small as inverse dimensionless
conductance of the normal metal $1/G_{m}$.
Instead, the magnitude of the effect is proportional to the inverse
conductance of the superconductor in the normal state $1/G_{s}$.
Under the condition $G_{m}\gg G_{s}$, which we assume in the present 
paper, this effect  
dominates the real energy relaxation. 

The remainder of the paper is organized as follows. In
Sec.~\ref{sec:2}, we  present a phenomenological derivation of our
main results. In the same section we suggest an independent experiment
to test our theory. Section ~\ref{sec:3} is devoted to the rigorous
analysis of the  the tunneling density of states 
under non-equilibrium conditions.
Our findings are summarized in Conclusions. 
Some mathematical details are relegated to Appendices.

\section{Qualitative discussion}
\label{sec:2}

The purpose of this Section is to describe  qualitatively how the
the non-equilibrium fluctuations affect the density of states of
the superconductor. 
We need first to classify the collective excitations, 
which are present in the system, 
and to understand how  non-equilibrium conditions influence them.

We start by recalling the basic physics of phase fluctuations in
superconductors, then consider their coupling to currents in the
metallic layer.  The electric current in the normal metal is
accompanied by shot-noise. As we show below, 
this noise gives rise to the classical phase fluctuations. 
Finally, we demonstrate 
that the enhanced fluctuations dramatically affect the BCS
density of states.
In particular, they lead to a non-zero density inside the BCS gap.  
This Section is concluded by
suggesting an independent experiment  to test our theory.

\subsection{Collective modes in N-S sandwich}
\label{col}

Consider a superconducting film at zero temperature. 
It is well known, 
that all of the excitations with the energy smaller than the
superconducting gap $\Delta$ 
are associated with the phase $\theta$ of the order parameter. 
The time evolution of this phase is
governed by hydrodynamic equations, which 
in the absence of external magnetic fields 
can be written as\cite{Tinkham}

\begin{mathletters}
\begin{eqnarray}
&&
\dot n_s + \frac{1}{2e}\mbox{\boldmath $\nabla\cdot j$}_s = 0,
\label{nep}\\
&&
\nonumber\\
&&
\mbox{\boldmath$ j$}_s = -e \pi\hbar D_s \nu_s \Delta \mbox{\boldmath
$\nabla$}\theta,
\label{cur}\\
&&
\nonumber\\
&&
\hbar\dot \theta = 2\left(e\varphi +  {{n_s}\over{\nu_s}}\right),
\label{jos}
\end{eqnarray}
\label{leq}
\end{mathletters}

\noindent
where $n_s$ is the perturbation of
the carrier density in the superconductor,
 $\nu_s$ is the thermodynamic density of states 
per unit area in the superconductor, and
$\mbox{\boldmath $j$}_s$ is the supercurrent. 
Equation~(\ref{nep}) is the continuity relation. 
We wrote the London equation (\ref{cur}) 
for a dirty superconductor 
and expressed the superfluid density
through the diffusion coefficient $D_s$ 
in the normal state of the superconductor. 
Equation (\ref{jos}) is the conventional Josephson
relation between the electrochemical potential 
and the phase of the order parameter $\theta$. 
Finally, the electrostatic potential $\varphi$ is connected to the
density variation by the Coulomb law

\begin{equation}
e\varphi = \int dr'V(r-r') n_s(r'),\quad V(r)=\frac{e^2}{r}.
\label{ep}
\end{equation}

Performing the Fourier transform 
of Eqs.\ (\ref{leq}) and \ (\ref{ep}),
we obtain the dispersion relation for the collective mode

\begin{equation}
\omega^2 = {\pi\over{\hbar}}
\Delta D_s Q^2\left[ 1+\nu_s V(Q)\right], \quad V(Q)=\frac{2\pi e^2}{Q}
\label{dr}
\end{equation}

\noindent
It corresponds  to the usual 2D
plasmon with dispersion $\omega\simeq \sqrt{Q}$ which has little
effect on the behavior of the system because of its small density of
states at low frequencies.

When a layer of the normal metal is brought nearby, 
$V(Q)$ gets screened, 
and the dispersion relation Eq.\ (\ref{dr}) becomes linear. 
To see this, one has to include the normal currents into the set of 
the hydrodynamic equations (\ref{leq})~--~(\ref{ep}). The equation for
the scalar potential (\ref{ep}) is modified to

\begin{equation}
e\varphi = \int dr'V(r-r')\left[ n_s(r')+n_m(r') \right] .
\label{epn}
\end{equation}

\noindent
Here we neglected for simplicity  the thickness of the isolating layer
between the normal metal and the superconductor assuming that
$d \ll 1/Q$.
All the results are insensitive to this assumption (see
Sec.~\ref{results}). The density of carriers $n_m$ in the normal metal
is governed by the continuity equation and the Ohm's law:

\begin{mathletters}
\begin{eqnarray}
&&
\dot n_m + \frac{1}{e}\mbox{\boldmath $\nabla\cdot j$}_m = 0,
\label{nepm}\\
&&
\nonumber\\
&&
\mbox{\boldmath$ j$}_m = -\sigma_m \mbox{\boldmath $\nabla$} \varphi -
D_m \mbox{\boldmath $\nabla$} n_m.
\label{curm}
\end{eqnarray}
\label{meq}
\end{mathletters}

The charge in the normal metal is redistributed by the electric field
according to Eqs.~(\ref{meq}). 
We evaluate $n_m$ from these equations,
substitute the result into Eq.~(\ref{epn}), 
and find, that the potential becomes dynamically screened. 
At frequencies much smaller, than the plasmon frequency in the normal
layer,
${\omega_p \simeq \nu_m V_0(Q)D_m Q^2}$, 
the screened potential takes the form

\begin{equation}
V(Q,\omega ) = {1\over{\nu_m}}
{{-i\omega + D_m Q^2}\over{D_m Q^2}}.
\label{vq}
\end{equation}

\noindent
We now substitute Eq.\ (\ref{vq}) into the dispersion law
of the collective mode Eq.\ (\ref{dr}) and find

\begin{mathletters}
\begin{eqnarray}
&&
\omega_{ph} = \omega_{ph}' - i\omega_{ph}'',
\\
&&
\nonumber\\
&&
\omega_{ph}'=Q\left({{\pi\Delta D_s}\over{\hbar}}\right)^{1/2}
\left(1+{{\nu_s}\over{\nu_m}}\right)^{1/2}
\label{or}
\\
&&
\nonumber\\
&&
\omega_{ph}''= {\pi\over{2}} \left({{\nu_s D_s}\over{\nu_m D_m}}\right)
{\Delta\over{\hbar}}.
\label{oi}
\end{eqnarray}
\label{sp}
\end{mathletters}

\noindent
Equations (\ref{oi}) -- (\ref{or}) are valid provided that
 $\omega_{ph}' > \omega_{ph}''$. This condition is satisfied
 already at small frequencies $\hbar\omega \simeq \hbar\omega_{ph}'
 \simeq \Delta (G_{s}/G_{m}) \ll \Delta$. The lifetime of this mode
is finite due to the interaction with the relaxation mode in the 
normal metal.
 
Since $\omega_{ph}' > \omega_{ph}''$, Eqs.~(\ref{or}) and (\ref{oi}) describe 
a well pronounced collective mode. We will call this mode ``phason''.  
According to Eq.\ (\ref{or}), the phason has linear dispersion.
This is due to the fact that the currents
in the normal metal flow along the interface 
in direction opposite to the currents in the
superconducting layer, 
so the total charge accumulation does not occur.
The physics of this mode is essentially similar to the well known
Schmid - Sch\"on\cite{schm} mode 
in the vicinity of the critical temperature or
to the second sound in superfluids\cite{secsound}. 
The only difference is that the normal excitations  are not thermally
activated in the superconductor itself 
but rather exist in the normal metallic layer
close to the superconductor. 
This,  however, has little  consequence on the charge dynamics.

\subsection{Phase fluctuations due to the current in the normal layer}

We have seen that a presence of the metal gives rise to
the new collective mode, the phason. 
This mode corresponds to the oscillating electric currents 
flowing in the opposite directions in the
metallic and the superconducting layers. 

Now let us consider what happens,
when a current is driven in the normal layer. 
The average currents in the metal
are accompanied by the fluctuations known as the shot noise.
Since the currents in the metal are coupled to those in the
superconductor, it is natural to expect that 
in the superconductor the fluctuating currents appear as well, 
and consequently, the  phasons are generated. 
In other words, a {\em dc}-current in the metallic layer should
enhance phase fluctuations in the superconductor. 

To include these fluctuations in our
description of the {N-S} sandwich we  add Langevin sources $\delta
\mbox{\boldmath$ j$}_l$ to the current in the normal metal.  
Equation (\ref{curm}) takes the form

\begin{equation}
\mbox{\boldmath$ j$}_m = -\sigma_m \mbox{\boldmath $\nabla$} \varphi -
D_m \mbox{\boldmath $\nabla$} n_m + \delta \mbox{\boldmath$ j$}_l.
\label{cml}
\end{equation}

\noindent
The fluctuations $\delta \mbox{\boldmath$j$}_l$ 
are described by their correlator 
$\langle \delta j_l^\alpha \delta j_l^\beta \rangle_{\omega,Q}$.
Provided the frequency $\omega$ is
much less than the applied voltage $\omega \ll eU/\hbar$ 
and the energy relaxation is negligible,
this correlator can be written as \cite{shot}

\begin{equation}
\langle \delta j_l^\alpha \delta j_l^\beta \rangle_{\omega,Q} =
\delta_{\alpha\beta}\sigma_m eU \propto e\langle j_m \rangle.
\label{cc}
\end{equation}

The superconducting phase $\theta$ 
in the presence of the current fluctuations $\delta
\mbox{\boldmath$j$}_l$
can be determined from 
the system of equations  (\ref{leq}),  \ (\ref{epn}), 
\ (\ref{nepm}) and \ (\ref{cml}). In addition, we use the Einstein
relation $\sigma_m = e^2 \nu_m D_m$. As a result, we can 
present the phase fluctuation $\delta\theta$ as

\begin{equation}
\delta\theta \propto -i{\delta {\mbox{\boldmath$j$}_l \cdot
\mbox{\boldmath$Q$}}\over{e\hbar\nu_m D_m Q^2}}
{\omega\over{\omega^2-\omega_{ph}^2(Q)}}
\label{theta}
\end{equation}

\noindent
where $\omega_{ph}^2(Q)$ is the phason dispersion.
In Eq.~(\ref{theta}) and all the subsequent formulas, 
we have not specified an inessential numerical prefactor,
which will be found in the next section.
Using the correlator Eq.\ (\ref{cc}) we obtain

\begin{equation}
\langle \delta\theta^2 \rangle_{\omega, Q} = {{eU}\over{\hbar^2\nu_m D_m
Q^2}}
{{\omega^2}\over{|\omega^2-\omega_{ph}^2(Q)|^2}}.
\label{pc}
\end{equation}

Therefore, the correlator of the phase fluctuations 
$\langle \theta^2 \rangle_{\omega, Q}$
has a well pronounced phason pole and 
is proportional to the applied voltage $U$.

In what follows, we will need the single point correlator of the phase
fluctuations, {\em  i.e.}, the integral of Eq.~(\ref{pc}) 
over the momentum $Q$.
The main contribution to this integral comes from the pole,
which corresponds to the resonant excitation of the phason 
by the current fluctuations in the normal layer.
The logarithmic divergence at $Q=0$ in Eq.\ (\ref{pc}) 
is not important and,
as we show in Sec.~\ref{zm}, disappears due to the gauge invariance. 
Since the linewidth of the phason decreases 
with the increase of the diffusion coefficient
$D_m$, see Eq.~(\ref{oi}), the large factor $D_m$ in the denominator of
Eq.~(\ref{pc}) is canceled. 
As a result,  the single point correlator 
of the phase fluctuations $\langle \delta\theta^2 \rangle_{\omega}$
does not depend on the parameters of the metallic layer

\begin{equation}
\langle \delta\theta^2 \rangle_{\omega} =
\int d^2Q \langle \delta\theta^2 \rangle_{\omega, Q}
= {{eU}\over{\Delta}} {1\over{G_s|\omega|}}.
\label{pco}
\end{equation}

\noindent
Equation (\ref{pco}) is valid, provided 
$G_s \Delta /G_m \ll \hbar\omega \lesssim eU$, where 
$G_{s,m}$ denote dimensionless conductances 
of the superconducting (in the normal state) and normal
layers respectively:
$G_{s,m} =2\pi\hbar \sigma_{s,m} /e^2 = 2\pi\hbar\nu_{s,m} D_{s,m}$,
these are the conductivities
measured in units of $e^2/{2\pi\hbar} = 1/ (25.8 K\Omega)$.

\subsection{Effect of the phase fluctuations
on the tunneling DOS of the superconductor}

We have found that in the presence of the normal layer the 
phase fluctuations in the superconductor are large 
due to the resonant excitation of the phasons.
Now we are interested in the effect  
of these fluctuations on a measurable quantity,
{\em e.g.} on the tunneling conductance, $G_T = \partial I/\partial V$ 
of the junction, with
$I$ and $V$ being the current and the voltage across the junction
respectively.

In the lowest order in the tunneling amplitude, 
the tunneling conductance $G_T$ is 
determined by the density of
states of the superconductor Eq.~(\ref{Gt}). 
The density $\rho(\epsilon)$ depends on 
single particle excitation energies.
In the absence of fluctuations the excitation energy 
in the superconductor is given
by the usual BCS expression $E=\sqrt{\xi^2+\Delta^2}$, 
where $\xi$ is the energy of the orbital state
counted from the Fermi level. 
The energy $E$ can not be smaller than $\Delta$. 
This prevents electrons (or holes) from the metal
with energies smaller than $\Delta$ from tunneling into the superconductor.

However, in the presence of the phase fluctuations, 
it becomes
possible for an electron with the energy smaller than $\Delta$
to tunnel and then to absorb a phason 
to compensate for the energy deficit. 
As a result, the density of states turns out to be finite 
even inside the gap $E < \Delta$.

To describe this effect of the phason assisted tunneling 
more quantitatively, we first calculate the density of states
in the superconductor in the presence of homogeneous phase fluctuations. 
In this case, due to the orthogonality of the orbital wave functions,
all the transitions are confined to the same orbital 
(characterized by some orbital energy $\xi$).
The problem simplifies since within the single orbital we have to consider
only
four states
(one orbital can be occupied by no more than two
electrons): $\psi_0$ - empty orbital, $\psi_2$ filled orbital,
$\psi_\uparrow$ and $\psi_\downarrow$ - singlets.
States  $\psi_0$ and  $\psi_2$ are coupled to each other
due to the exchange with the condensate, whereas singlets are not.
The resulting Schr\"odinger equations are

\begin{mathletters}
\begin{eqnarray}
&&
i\hbar \dot \psi_0 = \Delta e^{2i\theta(t)} \psi_2,\\
&&
\nonumber\\
&&
i\hbar \dot \psi_2 = 2 \xi \psi_2 + \Delta e^{-2i\theta(t)} \psi_0,\\
&&
\nonumber\\
&&
i\hbar \dot \psi_{\uparrow, \downarrow} = \xi \psi_{\uparrow, \downarrow}.
\end{eqnarray}
\label{sch}
\end{mathletters}

\noindent
At frequencies smaller than $eU/\hbar$ the occupation number of
phasons is large $\simeq eU/\hbar\omega$, and, therefore, 
$\theta$ can be treated as a classical variable.
 
Phase $\theta$  changes with the characteristic frequency of the order 
of  $eU/\hbar \ll \Delta/\hbar$. 
Hence, Eqs.~(\ref{sch}) can be solved in the adiabatic approximation.
The time-dependent ground state of this four-level system is given by

\begin{equation}
\Psi_{GS}(t) = \left [u \psi_0 + v \psi_2 e^{2i\theta(t)} \right ]
e^{-{i\over{\hbar}}\int^t dt_1 E_{GS}(t_1)},
\label{gs}
\end{equation}

\noindent
where  $u$ and $v$ are the usual coherence factors, $u = \cos\alpha$, 
$v=\sin\alpha$ and
$\tan 2\alpha = - \Delta/\tilde\xi$,and

\begin{eqnarray}
E_{GS}(t) = \tilde\xi -
\sqrt{\tilde\xi^2+\Delta^2},
\label{gse}\\
\tilde\xi =\xi - \hbar\dot\theta(t).
\nonumber
\end{eqnarray}

\noindent

So far we discussed an isolated superconductor. 
Let us now consider tunneling of an electron 
with the energy $\epsilon$ from
the normal metal into the superconductor.
Since the initial state has the energy $E_{GS} +\epsilon$ and
the final state is one of the states 
$\psi_{\uparrow(\downarrow)}$, each having energy $\xi$, 
the transition amplitude of this process  can be estimated as

\begin{equation}
{\cal A}(t) = \dots \int\limits^t_0 dt'
\exp \left({i\over{\hbar}} \int\limits^{t'}_0 dt''
[\epsilon +E_{GS}(t'')-\xi] \right).
\label{A}
\end{equation}

\noindent
The prefactor in Eq.~(\ref{A}) denoted by $\dots$ includes 
the tunneling matrix element and the coherence factors. 
This prefactor is not important for present discussion.
The transition rate is given by

\begin{equation}
{1\over{\tau_{tunn}}} \propto
\lim_{t\to \infty } {1\over{t}} \langle |{\cal A}(t)|^2 \rangle_\theta
\label{tun}
\end{equation}

\noindent
where $\langle\dots \rangle_\theta$ stands for the averaging over the
fluctuations of the phase $\theta$.
 
We can substitute the amplitude Eq.~(\ref{A})
into the tunneling rate Eq.~(\ref{tun}) 
and use the explicit form of $E_{GS}(t)$ given by Eq.~(\ref{gse}). 
In the exponent we have 
$\epsilon - \sqrt{\tilde\xi^2+\Delta^2} + \hbar\dot\theta$.

The density of states in the superconductor $\rho(\epsilon)$ 
is proportional to the tunneling rate ${1/\tau_{tunn}}$.
Evaluating the time integral in the exponent,
we  obtain
\begin{equation}
\rho(\epsilon)=\lim_{t\to \infty } {1\over{2\pi t}}
\langle\left|\int\limits^t_0 dt'
e^{i[\int_{0}^{t'}(\epsilon - \sqrt{\tilde\xi^2+\Delta^2}){dt''\over{\hbar}}
+ \theta (t')-\theta (0) ]}\right|^2\rangle_{\theta}.
\end{equation}

\noindent
The absolute value square of the time integral 
can be expressed in terms of the double
integral over $t'$ and $t''$. 
It is convenient to change variables to the sum 
and difference of $t'\pm t''$.
The exponent depends only on the time difference.
Indeed, the averaging over fluctuations of $\theta$ 
can be performed independently and the average 
functions depend on the time difference only. 
The integral over $t'+t''$ then yields $t$, 
which cancels with the denominator in the prefactor. 
In the remaining integral over the difference $t'-t''$ 
the upper limit can now be taken to infinity. 
The last step is to  sum over all
orbitals, since the electron can tunnel to any of them.
This amounts to integration over 
$\xi$ ($\sum\limits_{orb} \rightarrow \nu_s \int d\xi$).
This integral is independent from the fluctuations of 
$\theta$, since they produce shift of $\xi$ which is irrelevant 
because of the integration over $\xi$,
and yields the BCS density of states $\rho_0(\epsilon)$ 
Fourier transformed to the time domain:
\[
\sum_{orb}e^{i\int_{0}^{t}(\epsilon - 
\sqrt{\tilde\xi^2+\Delta^2}){dt'\over{\hbar}}}
=\int d\epsilon \rho_0(\epsilon) e^{i\epsilon t}.
\]
Thus, we find

\begin{equation}
\rho(\epsilon)=\int\limits^\infty_{-\infty} dt \int\limits^\infty_0
{{d\epsilon'
}\over{2\pi\hbar}}
\rho_0(\epsilon') e^{{i\over{\hbar}}(\epsilon - \epsilon')t}
\langle e^{i[\theta (t)-\theta (0)]}\rangle_\theta.
\label{ds}
\end{equation}

Therefore, in the presence of the phase fluctuations 
the density of states in the superconductor becomes modified 
by the factor, which describes the phason
assisted tunneling from the normal layer. 
To evaluate the average we use the
phase correlator Eq.~(\ref{pco}). 
It is not correct, since here we considered
only the homogeneous fluctuations of the phase, 
while the correlator Eq.~(\ref{pco})
includes also inhomogeneous fluctuations. 
However, the effect of the inhomogeneity
can be estimated as $\hbar D_s Q^2 /\Delta$, 
while the main contribution comes from the phason
mode with $D_s Q^2 \sim \hbar\omega^2/\Delta$,  
Therefore the correction is
$\sim\hbar^2\omega^2/\Delta^2$, and should be neglected because
the corrections of the same order were already omitted
within the adiabatic approximation.

Averaging the phase factor with the help of 
Eq.~(\ref{pco}), we find

\begin{eqnarray}
\rho(\epsilon)\approx &&
\int\limits^\infty_{-\infty} dt \int\limits^\infty_0
{{d\epsilon'}\over{2\pi\hbar}}
\rho_0(\epsilon') e^{{i\over{\hbar}}(\epsilon - \epsilon')t} \nonumber\\
&&
\nonumber\\
&&
\exp \left[ -{{eU}\over{\Delta G_s}}\int\limits_{-eU}^{eU}{{d\omega}\over
{|\omega|}}
\left( e^{i\omega t} - 1 \right) \right] .
\label{dsu}
\end{eqnarray}

\noindent
The integration is cut off at $\omega = eU$, 
because at $\omega > eU$ the classical description fails, 
and Eq.~(\ref{pco}) for the phase fluctuations is not valid.

In the energy interval $\Delta - eU \le \epsilon < \Delta$ 
one can expand the exponent to the first order, 
which corresponds to a single phason process. As a result,

\begin{eqnarray}
\rho(\epsilon) = 
\left( {{eU}\over{2 G_s\Delta }} \right)
\sqrt{\Delta\over{\Delta - \epsilon}},
\label{result}
\end{eqnarray}

\noindent
where numerical coefficient is the result of rigorous treatment of the
following section.
Equation~(\ref{result}) is the main qualitative result of the paper. 
It shows that in the
presence of the normal layer 
driven out of equilibrium by the applied voltage $U$, 
the superconductor no longer has a hard gap. 
Instead, there is a dip, and the tunneling density of states is suppressed
at $\epsilon < \Delta$ (see Fig.~\ref{2}).

{
\narrowtext
\begin{figure}[ht]
\vspace{0.2 cm}
\epsfxsize=7.7cm
\centerline{\epsfbox{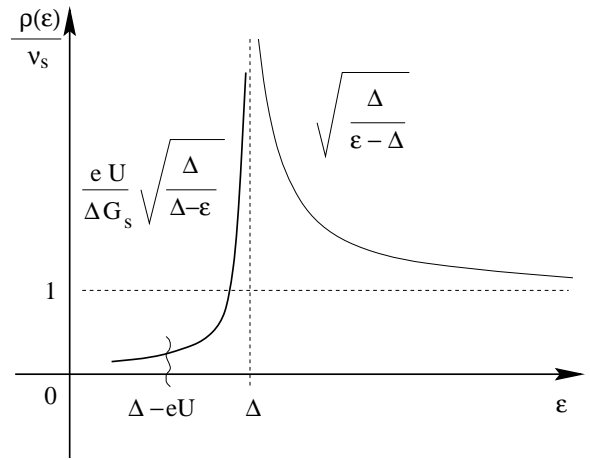}}
\vspace{0.2cm}
\caption{Sketch of the tunneling density of states 
$\protect\rho(\protect\epsilon)$
the superconductor modified by the current fluctuations in the normal
metal 
driven out of
equilibrium by the applied voltage $U$.}
\label{2}
\end{figure}
}

At energies smaller than  $\Delta - eU$ the density of states 
differs from zero due to multi-phason processes. 
It means that the exponent in Eq.~(\ref{dsu}) 
should be expanded to higher orders. 
However, to describe these multi-phason processes, we need 
better understanding of the nature of the frequency cut off. 
Also, we have so far neglected the quantum part 
of the phase fluctuations, $\omega>eU$.  
These issues are addressed in the technical part of the paper,
Sec.~\ref{sec:3}
where we perform a rigorous calculation 
of the tunneling, based on Keldysh diagrammatic technique.

\subsection{Is the effect of phason assisted tunneling independently
observable?
}
\label{sec:expt}

We have shown that the phason assisted tunneling 
manifests itself in the tail in the density of states 
within superconducting gap. 
When interpreting the experimental data, 
this effect could be confused with the broadening  of the distribution
function
$f_M(\epsilon)$ in the normal metal [see Eq.~(\ref{Gt})].
We believe that such a misinterpretation 
of the experimental data was made in Ref. \onlinecite{exp1}.

We should warn the reader that Eq.~(\ref{result}) as well as the
subsequent rigorous calculation of the tunneling current
[see Eq.~(\ref{cond-res}) for the result] were obtained for 
the simplest model of the N-S junction, namely the 2D sandwich. 
The spectrum of
collective modes is sensitive to the geometry of the
system, therefore, our results are  not expected to describe the experiment
of Ref.~\onlinecite{exp1} in detail. However, we have
presented a strong  evidence of the existence of the microscopic
mechanism responsible for the change of 
the shape of the tunneling 
conductance, which is different from
the trivial broadening of the distribution function
in the normal metal.
In this paper we do not intend to evaluate the effect of 
phason assisted tunneling in various possible experimental
realizations of the N-S junction. Instead, we suggest an independent
measurement, that can distinguish between the two mechanisms.
  
{
\narrowtext
\begin{figure}[ht]
\vspace{0.2 cm}
\epsfxsize=7.7cm
\centerline{\epsfbox{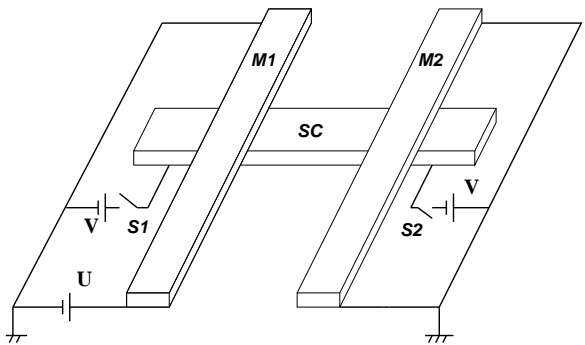}}
\vspace{0.2cm}
\caption{Scheme of the experiment for observation the effect of
non-equilibrium phase fluctuations
on the density of states. Switch ``S1'' corresponds to the experiment of
Ref.~\protect\onlinecite{exp1}. Switch ``S2'' corresponds to the
measurement
of the density of states affected by the current fluctuations in
non-equilibrium
electrode ``M1'' but not convoluted with its distribution function.}
\label{3}
\end{figure}
}
 
The suggested experimental setup is shown on Fig~\ref{3}. 
This scheme differs from the experiment  of  Ref.~\onlinecite{exp1}
by adding another normal electrode ``M2''. 
The electrode ``M1'' is driven out of
equilibrium by the applied voltage $U$. 
That affects the density of states in the
superconductor (``SC'') due to phase fluctuations as well as broadens 
the distribution function in
the metal ``M1'' itself due to energy relaxation. 
We suggest to measure the tunneling conductance between
the superconductor and the second electrode ``M2''. 
Since this electrode is in equilibrium, 
its distribution function is broadened by temperature only.  
Such measurement will only show the modification 
of the density of states of the superconductor by
the current fluctuations in "M1" (shown on Fig.~\ref{2}). 
Comparing the two measurements, one should be able to determine which
effect dominates the tunneling between the superconductor 
and the non-equilibrium normal metal.

\section{Theory of the interaction effects in N-S junctions.}
\label{sec:3}

In this Section we set $\hbar = 1$ in all intermediate expressions.

\subsection{Tunneling current}

In this section we use
Keldysh \cite{Keldysh,Larkin77} diagrammatic technique to evaluate 
the tunneling current  $I$ 
between the superconductor and the normal metal.
In order to take superconductivity into account properly,
we need Green functions
${\cal G}$ to be matrices in the space $K\otimes N$, 
which is the direct product of Keldysh space $K$ 
and Gor'kov-Nambu\cite{nam} space $N$.
We choose the basis in the Keldysh space, 
for which electronic Green's functions are \cite{Larkin77}

\begin{eqnarray}
{\cal G} =
\pmatrix
{
{\cal G}^R  & {\cal G}^K  \cr
0           & {\cal G}^A  \cr
}_K ,
\label{kgf}
\end{eqnarray}

\noindent
Let $[\; ]_-$ and $[\; ]_+$ denote a commutator  
and anticommutator correspondingly. 
The three components of the Green's function ${\cal G}_{\alpha\beta}$, 
where $\alpha$ and $\beta$ can be either 
$M$ (metal) or $S$ (superconductor), 
are defined as

\begin{mathletters}
\begin{eqnarray}
&&
{\cal G}_{\alpha\beta}^K(t_1, t_2) =
- i \langle \left [\Psi_\alpha(t_1)
\otimes\Psi_\beta^\dagger(t_2)\right ]_-\rangle,\\
&&
\nonumber\\
&&
{\cal G}_{\alpha\beta}^R(1, 2) =
- i \eta(t_1-t_2) \langle \left [\Psi_\alpha(1)
\otimes\Psi_\beta^\dagger(2)\right ]_+\rangle,\\
&&
\nonumber\\
&&
{\cal G}_{\alpha\beta}^A(1, 2) =
i \eta(t_2-t_1) \langle \left [\Psi_\alpha(1)
\otimes\Psi_\beta^\dagger(2)\right ]_+\rangle,
\end{eqnarray}
\label{msfunction-def}
\end{mathletters}

\noindent
where we used a short-hand notation $n = (\vec r_n, t_n)$ for
$n = 1,2$.  
In Eqs.~(\ref{msfunction-def})
$\eta(x)$ is Heaviside step function, 
$\Psi$ are Nambu \cite{nam} spinors

\begin{eqnarray}
\Psi =
\pmatrix
{
{  \psi_\downarrow }\cr
{  \psi^\dagger_\uparrow}\cr
}
; \Psi^\dagger = \Big ( \psi^\dagger_\downarrow ; - \psi_\uparrow \Big ),
\end{eqnarray}

\noindent
and $\psi_\downarrow (n)$, $\psi_\uparrow (n)$ 
are the fermionic operators
in the Heisenberg representation. 
Since we are using the Keldysh technique, $\langle \dots \rangle$
denotes the average with arbitrary distribution function.

We start our calculation of the tunneling current 
$I$ between the superconductor and the normal metal 
with the general expression

\begin{equation}
I = {e\over{2}} T_0 {\bf Tr} 
\big [ \sigma_2 (\cal G_{MS}- \cal G_{SM}) \big ],
\label{current-def}
\end{equation}

\noindent
Within our choice of the basis [Eq.~\ref{kgf} the  current vertex 
in Eq.\ (\ref{current-def}) is Pauli matrix $\sigma_2$ 
in Keldysh space

\begin{eqnarray}
\sigma_2 =
\pmatrix
{
0 & 1  \cr
1 & 0  \cr
}_K
\otimes
\pmatrix
{
1 & 0  \cr
0 & 1  \cr
}_N
\label{s22}
\end{eqnarray}

\noindent
${\bf Tr}$ in Eq.\ (\ref{current-def}) means trace in the  $K\otimes N$
space and also assumes that $\vec r_1 = \vec r_2 = \vec r$, 
where $\vec r$ corresponds to the contact point. 
Below we often ommit the spatial coordinates of the Green functions
and write explicitly only temporal coordinates. 
It always assumes that $\vec r_1 = \vec r_2 = \vec r$. 

In the first order in the tunneling amplitude $T_0$ the Green's 
functions can be calculated as

\begin{equation}
{\bf {\cal G}}_{MS} = -\pi^2\nu_m\nu_s T_0
\int dt^\prime {\bf g}_M(t, t^\prime) e^{-ieVt^\prime} 
{\bf g}_S(t^\prime, t),
\label{gf-firstorder}
\end{equation}

\noindent
where $V$ is the applied voltage and ${\bf g}_M$ and ${\bf g}_S$ are 
the Green's
functions in the normal metal and in the superconductor, respectively, defined
as

\begin{equation}
{\bf g}_{\alpha}(t_1, t_2) = {i\over{\pi\nu_{\alpha}}} 
{\cal G}_{\alpha\alpha}(\vec r, \vec r; t_1, t_2).
\label{qgf}
\end{equation}

\noindent
In what follows the semiclassical Green's functions in the $K\otimes N$ 
space will be denoted by boldfaced symbols, 
while the $2\times 2$ matrices in the Nambu space will be denoted by a hat.

In the normal metal the semiclassical Green's function 
${\bf g}_{M}$ satisfies the Usadel equation

\begin{eqnarray}
- D_m\nabla^c_R &&( {\bf g}_M \circ \nabla^c_R {\bf g}_M)
\nonumber\\
&&
\nonumber\\
&&
+ {\bf \tau}^3 {\partial {\bf g}_M \over{\partial t_1}} +
 {\partial {\bf g}_M \over{\partial t_2}}{\bf \tau}^3
- i [ {\bf \Phi}, {\bf g}_M ]_t = 0,
\label{usm}
\end{eqnarray}

\noindent
where the matrix $\tau^3$ is

\begin{eqnarray}
\tau^3 =
\pmatrix
{
1  & 0  \cr
0  & 1  \cr
}_K \otimes
\pmatrix
{
1 &  0  \cr
0 & -1  \cr
}_{N},
\end{eqnarray}

\noindent
the commutator designates

\begin{equation}
[{\bf \Phi}, {\bf g}_M ]_t = {\bf \Phi} (t_1) {\bf g}_M(t_1, t_2)
- {\bf g}_M(t_1, t_2) {\bf \Phi} (t_2),
\label{tcom}
\end{equation}

\noindent
and

\begin{equation}
{\bf g}_M \circ {\bf g}_M (t_1, t_2) = \int dt_3 {\bf g}_M (t_1, t_3) 
{\bf g}_M(t_3, t_2).
\label{circ}
\end{equation}

\noindent
The scalar potential is also a matrix

\begin{eqnarray}
{\bf \Phi} =
\pmatrix
{
\varphi_+  & \varphi_-  \cr
\varphi_-  & \varphi_+  \cr
}_K \otimes
\pmatrix
{
1 &  0  \cr
0 &  1  \cr
}_{N}.
\end{eqnarray}

\noindent
While the uniform scalar potential can be gauged out 
and does not affect the behaviour of the system, 
it can be shown \cite{antn} that
the slow spatial variation of $\varphi$ can be
taken into acount by means of linear functional $K[\varphi]$ 
so that the resulting Green's function can be written as

\begin{equation}
{\bf g}_M(t_1, t_2) = e^{iK[\varphi](t_1)\tau^3} \left ({\bf g}_{M0}
+{\bf g}_{M1} \right) e^{-iK[\varphi](t_2)\tau^3},
\label{mfunc}
\end{equation}

\noindent
where ${\bf g}_{M0}(t_1 - t_2)$ is the 
noninteracting Green's function, 
while ${\bf g}_{M1}(t_1 , t_2)$
contains all of the corrections not captured 
by the transformation Eq.\ (\ref{mfunc}).

The functional $K[\varphi]$ can be specifically chosen 
in such a way, that in equilibrium the
correction ${\bf g}_{M1}(t_1 , t_2)$ 
is porportional to the square of the gradient
${{\bf g}_{M1} \sim (\nabla K[\varphi])^2}$. 
When the system is out of equilibrium, 
${\bf g}_{M1}(t_1 , t_2)$ also acquires a term
proportional to the deviation 
of the distribution function  $f_M(\epsilon)$
from  equlibrium one $f_{M, eq}(\epsilon)$, i.e.
${\bf g}_{M1} \sim \Delta K[\varphi] 
(f_M(\epsilon) - f_{M, eq}(\epsilon))$. 
Such functional
is given by \cite{antn} (see also Appendix A)

\begin{mathletters}
\begin{eqnarray}
&&
{\; \; \; \; \; \; \; \; \; \; \; \; \; \; \; \; \; \; } \pmatrix
{
K[\varphi]_+  \cr
K[\varphi]_-  \cr
}
= {\cal K}
\pmatrix
{
\varphi_+  \cr
\varphi_-  \cr
},\\
&&
\nonumber\\
&&
{\cal K} =
\pmatrix
{
(-i\omega + D_m Q^2)^{-1} & -2  {\rm Re}{{N(\omega)}
\over{-i\omega + D_m Q^2}} \cr
0                         & -(i\omega + D_m Q^2)^{-1} \cr
},
\end{eqnarray}
\label{Kfunc}
\end{mathletters}

\noindent
where $N(\omega)$, which can be named  
``bosonic distribution function''  is defined as

\begin{eqnarray}
N(\omega)=\int {{d\epsilon}\over{\omega}}
\Big [2f_M(\epsilon)&&f_M(\epsilon +\omega) - f_M(\epsilon)\nonumber\\
&&
\nonumber\\
&&
 - f_M(\epsilon +\omega) \Big ].
\label{fw}
\end{eqnarray}

\noindent
In this case the corrections ${\bf g}_{M1}(t_1 , t_2)$ 
in Eq.\ (\ref{mfunc}) come from the effects of energy relaxation 
and interaction localization corrections 
(the effects of weak localization were negleced 
from the very beginning, 
as discussed above). 
These corrections are inversely proportional
to the metal conductance $G_m$ (in 2D; in 1D the 
corrections are $\sim 1/ G_m(l_{ph})$ where
$l_{ph}$ is the dephasing length; for details see Ref. 4). 
Therefore, these corrections can also be neglected.
 
The retarded and advanced Green functions 
$g^{R,A}_{M0}$ in the time domain are delta functions

\begin{equation}
g_{M0}^R(t) = - g_{M0}^A(t) = \delta(t).
\label{mfuncR0}
\end{equation}

\noindent
The Keldysh function in the energy domain 
is related to the distribution
function by

\begin{equation}
g_{M0}^K(\epsilon) = 2(1-2f_M(\epsilon)).
\label{mfuncK0}
\end{equation}

Fluctuating electric fields
give rise to the fluctuations of the phase $\theta$ 
of the order parameter $\bar {\bf \Delta}$ in the superconductor .
The fluctuations of the absolute value of the order parameter $\Delta$ 
do not couple to the fluctuations of the electric field, 
and their spectrum has a gap $\sim\Delta$. 
Therefore they can be ignored. 
The superconducting Green's function ${\bf g}_S$ 
can be evaluated by solving the Usadel equation in the superconductor

\begin{eqnarray}
- D_s\nabla^c_R &&( {\bf g_S} \circ \nabla^c_R {\bf g}_S)
\nonumber\\
&&
\nonumber\\
&&
+ {\bf \tau}^3 {\partial {\bf g}_S \over{\partial t_1}} +
 {\partial {\bf g}_S \over{\partial t_2}}{\bf \tau}^3
- i [ ({\bf \bar\Delta} + {\bf \Phi}), {\bf g}_S ]_t = 0,
\label{useq}
\end{eqnarray}

\noindent
where

\begin{eqnarray}
{\bf \bar\Delta}_0 =
\pmatrix
{
0                      & \Delta e^{i\theta_K} \cr
-\Delta e^{-i\theta_K} & 0                    \cr
}_N,
\end{eqnarray}

\noindent
and

\begin{eqnarray}
\theta_K =
\pmatrix
{
\theta_+  & \theta_-  \cr
\theta_-  & \theta_+  \cr
}_K.
\end{eqnarray}

In Appendix B we show that at small frequencies ($\omega\sim V \ll \Delta$) 
the dominant effect is captured by the gauge transformation

\begin{equation}
{\bf g}_S(t_1, t_2) = e^{i\theta_K(t_1)\tau^3} 
{\bf g}_{S0} (t_1 - t_2) e^{-i\theta_K(t_2)\tau^3},
\label{sfunc}
\end{equation}

\noindent
where ${\bf g}_{S0}$ is the usual BCS Green's function.

We now substitute the Green's functions in the normal and 
superconducting layers Eqs.~(\ref{mfunc}) and
 (\ref{sfunc}) into Eq.~(\ref{gf-firstorder}) and 
substitute the result into the expression 
for the tunneling current Eq.~(\ref{current-def}).  
The next step is to average over the phase and Coulomb fluctuations. 
The current is then given by the integral

\begin{eqnarray}
I = {{G_0}\over{2}}\int dt^\prime && e^{-ieVt^\prime}
{\bf Tr} \Big \{ \sigma_2 \big [ {\bf g}_{M0}(t-t^\prime) 
\bar {\bf g}_S (t^\prime -t)
\nonumber\\
&&
\nonumber\\
&&
- \bar {\bf g}_S(t-t^\prime ) {\bf g}_{M0} (t^\prime -t) \big ] \Big \},
\label{current-av}
\end{eqnarray}

\noindent
where $G_0 = 2 \pi \nu_m \nu_s e^2 T_0^2$. 
We used the cyclic property of the trace and an obvious fact 
that the matrices
$\sigma_2$ and $\tau^3$ to obtain the averaged function

\begin{equation}
\bar {\bf g}_S(t_1-t_2) = \langle e^{i\phi(t_1)\tau^3}
{\bf g}_{S0}(t_1-t_2) e^{-i\phi(t_2)\tau^3} \rangle,
\label{sfunc-av}
\end{equation}

\noindent
where the fluctuation field equals to

\begin{equation}
\phi = \theta - K[\varphi].
\label{fluct.field}
\end{equation}

So far our analysis was quite general and
independent of the geometry of the sample. 
Even though we were considering tunneling through
point contacts,  Eqs.\ (\ref{current-av}) and \ (\ref{sfunc-av}) 
can be strightforwardly generalized 
for the wide contacts with large number of channels.

We now calculate the trace in Keldysh space 
and use the explicit form of the metallic
Green's functions Eq.\ (\ref{mfuncR0}) and Eq.\ (\ref{mfuncK0}) 
to evaluate the time integral.
The differential tunneling conductance $G_T = \partial I / \partial V$ 
can be written as

\begin{equation}
G_T = {{G_0}\over{2}}
\int {{d\epsilon}\over{2\pi}} {{\partial f_M}\over{\partial \epsilon}}
(\epsilon - eV)
{\rm Re } {\bf Tr} \big [ \bar {\bf g}_S^R(\epsilon) \tau^3 \big ].
\label{cond-def}
\end{equation}

The last step is to perform the average in Eq.\ (\ref{sfunc-av}). 
In the leading approximation in $1/G_m$ the fluctations are Gaussian, 
and we obtain (see Appendix C for details) for the function 
$\bar h_S^R(\epsilon) = 
{\bf Tr} \big [ \bar {\bf g}_S^R(\epsilon) \tau^3 \big]$

\begin{eqnarray}
\bar h_S^R(\epsilon) = \int\limits_0^\infty 
dt && e^{i\epsilon t} \Big [ h_0^{+-
}(t)
e^{2i \big 
( {\cal D}_{\phi\phi}^{+-}(t) - {\cal D}_{\phi\phi}^{+-}(0) \big )}
\nonumber\\
&&
\nonumber\\
&&
-  h_0^{-+}(t)
e^{2i \big ( {\cal D}_{\phi\phi}^{-+}(t) - {\cal D}_{\phi\phi}^{-+}(0)\big )} 
\Big ],
\label{avd-eps}
\end{eqnarray}

\noindent
where

\begin{equation}
h_0^{+-} (t>0) = - h_0^{-+} (-t) = {\Delta\over{\pi}} {\it K}_1(i\Delta t)
\label{bcspmt}
\end{equation}

\noindent
and ${\it K}_1$ is the modified Bessel function.
The fluctuation propagators are defined as

\begin{mathletters}
\begin{eqnarray}
&&
\langle \phi_+ \phi_+ \rangle = i {\cal D}_{\phi\phi}^K (t_1-t_2), \\
&&
\nonumber\\
&&
\langle \phi_+ \phi_- \rangle = i {\cal D}_{\phi\phi}^R (t_1-t_2), \\
&&
\nonumber\\
&&
\langle \phi_- \phi_+ \rangle = i {\cal D}_{\phi\phi}^A (t_1-t_2),\\
&&
\nonumber\\
&&
\langle \phi_- \phi_- \rangle = 0,
\end{eqnarray}
\label{props}
\end{mathletters}

\noindent
and 

\begin{equation}
{\cal D}_{\phi\phi}^{+-(-+)} = ({\cal D}_{\phi\phi}^K \pm 
{\cal D}_{\phi\phi}^R)/2.
\label{causal}
\end{equation}

The conductance can finally be expressed in terms of the function
$\bar h_S^R(\epsilon)$ Eq.~ (\ref{avd-eps}) as

\begin{equation}
G_T = {{G_0}\over{2}}
\int {{d\epsilon}\over{2\pi}} 
{{\partial f_M}\over{\partial \epsilon}}(\epsilon
- eV)
{\rm Re } \bar h_S^R(\epsilon).
\label{cond-av}
\end{equation}

So far our results are quite general. 
They describe the effect of the low frequency ($\omega <\Delta$)
collective mode on the tunneling conductance. 
In the following sections we evaluate the tunneling conductance 
Eq.~(\ref{cond-av}) for our particular setup. 
To do this we need the fluctuation propagators ${\cal D}$, 
which is evaluated in the following subsection.

\subsection{Propagators for low energy excitations.}
\label{zm}

Throughout this paper we are using the small parameter $1/G_m$. 
When the fluctuations propagators ${\cal D}_{\phi\phi}$ is calculated,
this parameter allows us
to restrict ourselves by ladder diagrams, i.e., to use RPA. 
These diagrams can be summed up by means of solving the Dyson-type equation

\begin{equation}
{\cal D} = {\cal D}_0 + {\cal D}_0 \Pi {\cal D}.
\label{dyeq}
\end{equation}

\noindent
In this equation the propagators ${\cal D}$ 
and the polarization operator $\Pi$ are
$8\times 8$ matrices defined in the $K\otimes MS\otimes\varphi\theta$ space. 
The $2\times 2$ space  $\varphi\theta$ describes 
the two fluctuation fields we have in the probl em. Since these
fluctuations take place in normal metal as well as in the superconductor, 
another $2\times 2$ space $MS$ is needed.

In the metal sector only the $\varphi\varphi$ polarization operator is present,
which yields the usual diffusion pole

\begin{equation}
(\Pi^{MM}_{\varphi\varphi})^R = {{\nu_m D_m Q^2} \over{D_m Q^2 - i\omega}},
\label{polm}
\end{equation}

\noindent
where $D_m$ is the diffusion constant in the normal metal.

In the superconductor there exist the phase 
as well as the Coulomb fluctuations. 
Corresponding polarization operators can be obtained
from the general relation  (ommiting the $SS$ superscript)

\begin{equation}
{1\over{\nu_s}}\Pi^{\mu\nu}{ik} =
\pi{{\delta}\over{\delta \varphi_k^\nu}}{\bf Tr}
\left[{\bf \Gamma}_i^\mu{\bf g}_
S\right]
+\delta_{i\varphi}\delta_{k\varphi},
\label{self}
\end{equation}

\noindent
where the indices $\mu$, $\nu$ can be $+$ or $-$.
The polarization operator $\Pi^{\mu\nu}_{ik}$ is the matrix 
in Keldysh space:

\begin{eqnarray}
\Pi_{ik} =
\pmatrix
{
\Pi_{ik}^{+-}  & \Pi_{ik}^{--}  \cr
0              & \Pi_{ik}^{-+}  \cr
}_K.
\end{eqnarray}

\noindent
Indices $i,k$ (which can be $\varphi$ or $\theta$) label 
the fluctuation parameter space.
The second term in Eq.\ (\ref{self}) is the ``anomalous'' 
contribution of the large $\epsilon$
region which is not captured by the Usadel equation 
(for details see, for example, Ref. 5).
The vertex ${\bf \Gamma}_i$ is a vector in the $\varphi\theta$ 
space (as indicated by index $i$)
and a matrix in $K\otimes N$ space given by

\begin{mathletters}
\begin{eqnarray}
&&
{\bf \Gamma}_\theta^\mu = i\Delta (\tau^\mu)_K \otimes
\pmatrix
{
0               & e^{2i\theta_K}  \cr
e^{-2i\theta_K}  & 0               \cr
}_N,
\\
&&
\nonumber\\
&&
{\bf \Gamma}_\varphi^\mu = i (\tau^\mu)_K \otimes {\bf 1}_N,
\end{eqnarray}
\end{mathletters}

\noindent
where $(\tau^+)_K = {\bf 1}_K$ and $(\tau^-)_K = (\sigma_2)_K$.

To proceed with this program and evaluate the polarization 
operators Eq.~ (\ref{self}) we need an
explicit form of the Green's function ${\bf g}_S$. 
The approximate solution Eq.~(\ref{sfunc-av})
which was obtained by the gauge transformation 
is not sufficient here, since the polarization operator is
a gauge invariant quantity. 
For this reason substitution of Eq.~ (\ref{sfunc-av}) 
into Eq.~ (\ref{self}) immediately gives zero. 
Therefore we have to take into account the gradient terms, 
which were neglected in Eq.~ (\ref{sfunc-av}). 
To do that we expand the Usadel equation 
Eq.~ (\ref{useq}) to the first order in fluctuation fields

\begin{eqnarray}
D_s Q^2 {\bf g}_{S0}(\epsilon_1) {\bf g}_{S1} && 
(\epsilon_1, \epsilon_2)\nonumber\\
&&
\nonumber\\
&&
- i [{\bf H}_0, {\bf g}_{S1}]_{\epsilon} = i 
[{\bf \delta H}, {\bf g}_{S0}]_{\epsilon},
\label{eq1}
\end{eqnarray}

\noindent
where ${\bf H}_0 = 1_K \otimes \hat H_0$, ${\bf \delta H} = {\bf \Phi}
+ 2i\Delta {\bf \Theta} +\delta\Delta$. The last term is irrelevant, 
since the fluctuations of the absolute value 
of the order parameter are gapped. 
Nambu space matrix $\hat H_0$ is given by

\begin{eqnarray}
\hat H_0 =
\pmatrix
{
\epsilon  & \Delta  \cr
-\Delta   & -\epsilon \cr
}_N.
\end{eqnarray}

\noindent
The gapless phase fluctuations have the following matrix structure
(see also Appendix B)

\begin{eqnarray}
{\bf \Theta} =
\pmatrix
{
\theta_+  & \theta_-  \cr
\theta_-  & \theta_+  \cr
}_K \otimes
\pmatrix
{
0 & 1  \cr
1 & 0  \cr
}_N.
\end{eqnarray}

The solution of the first order equation Eq.\ (\ref{eq1}) 
is the following Keldysh
matrix (omitting the subscript $S$ herefrom)

\begin{eqnarray}
{\bf g}_1 (\epsilon_1, \epsilon_2) =
\pmatrix
{
\hat g_1^R  & \hat g_1^K  \cr
\hat g_1^X  & \hat g_1^A  \cr
}_K.
\end{eqnarray}

\noindent
The anomalous function $\hat g^X$ is only present before averaging 
over the fluctuations.

\begin{eqnarray}
\hat g_1^X (\epsilon_1, \epsilon_2) = {i\over{ B_{-+}}} \Big [ &&
\hat g_0^A (\epsilon_1) \delta \hat H_- (\epsilon_1 - \epsilon_2) 
\hat g_0^R (\epsilon_2)
\nonumber\\
&&
\nonumber\\
&&
  - \delta \hat H_- (\epsilon_1 - \epsilon_2) \Big ] ,
\label{o1x}
\end{eqnarray}

\noindent
where

\begin{mathletters}
\begin{eqnarray}
&& B_{-+} =  D_sQ^2 - iS_-(\epsilon_1) -iS_+(\epsilon_2),
\\
&&
\nonumber\\
&&
S_\pm (\omega) = \sqrt{(\epsilon \pm i0)^2-\Delta^2}.
\end{eqnarray}
\label{denom}
\end{mathletters}

\noindent
The rest of the functions are given by

\begin{eqnarray}
\hat g_1^R&& = {i\over{B_{++}}}
\Big [ \hat g_0^R \delta \hat H_+ \hat g_0^R
  - \delta \hat H_+  +
\hat g_0^K \delta \hat H_- \hat g_0^R
\nonumber\\
&&
\nonumber\\
&&
+\hat g_1^X \tanh {\epsilon_1\over{2T}}
\left ( S_+(\epsilon_1)-S_-(\epsilon_1)\right ) \Big ],
\label{o1r}
\end{eqnarray}

\begin{eqnarray}
\hat g_1^A&& = {i\over{B_{--}}}
\Big [ \hat g_0^A \delta \hat H_+ \hat g_0^A
  - \delta \hat H_+  +
\hat g_0^A \delta \hat H_- \hat g_0^K
\nonumber\\
&&
\nonumber\\
&&
+\hat g_1^X \tanh {\epsilon_2\over{2T}}
\left ( S_+(\epsilon_2)-S_-(\epsilon_2)\right ) \Big ],
\label{o1a}
\end{eqnarray}

\noindent
and 

\begin{eqnarray}
\hat g_1^K&& = {i\over{B_{+-}}}
\Big [ \hat g_0^R \delta \hat H_+ \hat g_0^K +
\hat g_0^K \delta \hat H_+ \hat g_0^A
\nonumber\\
&&
\nonumber\\
&&
+\hat g_0^R \delta \hat H_- \hat g_0^A +
\hat g_0^K \delta \hat H_- \hat g_0^K
  - \delta \hat H_-
\nonumber\\
&&
\nonumber\\
&&
+\hat g_1^A \tanh {\epsilon_1\over{2T}}
\left ( S_+(\epsilon_1)-S_-(\epsilon_1)\right )
\nonumber\\
&&
\nonumber\\
&&
+\hat g_1^R \tanh {\epsilon_2\over{2T}}
\left ( S_+(\epsilon_2)-S_-(\epsilon_2)\right ) \Big ]
\label{o1k}
\end{eqnarray}

\noindent
in the obvious notation (compare with Eqs.\ (\ref{o1x}) and \ (\ref{denom})).

Using the first order solution we can calculate 
the polarization operators Eq.~(\ref{self})
for the phase and the Coulomb fluctuations in the superconductor. 
For the retarded component of the polarization operator we obtain

\begin{eqnarray}
(\Pi^{SS})^R = \nu_S
\pmatrix
{
1        & - i\omega            \cr
i\omega  & \omega^2 - \pi\Delta D_sQ^2  \cr
}_{\varphi\theta}.
\label{poop}
\end{eqnarray}

The solution of Dyson equation Eq.\ (\ref{dyeq}) is

\begin{equation}
{\cal D} = \big [ {\cal D}_0^{-1} - \Pi \big ]^{-1},
\label{dyeqinv}
\end{equation}

\noindent
where $\Pi$ is the polrization operator of the form

\begin{eqnarray}
\Pi =
\pmatrix
{
\Pi^{SS} & 0         \cr
0        & \Pi^{MM}  \cr
}_{MS}.
\end{eqnarray}

The unperturbed propagator ${\cal D}_0$ has non zero elements 
only in the Coulomb interaction channel

\begin{equation}
{\cal D}_0 = -V(Q)
\pmatrix
{
1       & e^{-Qd}  \cr
e^{-Qd} & 1        \cr
}_{SM}
\otimes 1_K \otimes
\pmatrix
{
1 & 0  \cr
0 & 0  \cr
}_{\varphi\theta},
\label{prin}
\end{equation}

\noindent
where the Coulomb interaction in two dimensions 
$V(Q) = 2\pi e^2 / Q$ and $d$ is the
thickness of the sample. 
This form of the unperturbed propagatoris model dependent. 
It corresponds to the particular setup
of the S-N sandwich, which we are discussing in this paper.

To obtain the retarded propagators we have 
to invert the retarded block of the $8 \times 8$ matrix. 
Since we  need only $\phi\phi$ propagator in Eq.\ (\ref{avd-eps}),
there are three elements of the remaining 
$4 \times 4$ matrix which are relevant:

\begin{mathletters}
\begin{eqnarray}
&&
{\cal D}^{\varphi\varphi}_M = {1\over{M}}
\left [ \left ( {1\over{\tilde V}} + \Pi_{\varphi\varphi}\right ) 
\Pi_{\theta\theta}
- \Pi_{\varphi\theta}\Pi_{\theta\varphi} \right ], \\
&&
\nonumber\\
&&
{\cal D}^{\varphi\theta}_{MS} = - {1\over{M}}
\left [ {{1-Qd}\over{\tilde V}} \Pi_{\theta\varphi} \right ], \\
&&
\nonumber\\
&&
{\cal D}^{\theta\theta}_S = {1\over{M}}
\Big [ \left ( {1\over{\tilde V}} + \Pi_{\varphi\varphi}\right )
\left ( {1\over{\tilde V}} + \Pi_M\right ) \nonumber\\
&&
\nonumber\\
&&
{\; \; \; \; \; \; \; \; \; \; \; \; \; \; \; \; \; }
-  {{(1-Qd)^2}\over{\tilde V^2}}\Big ],
\end{eqnarray}
\label{propsym}
\end{mathletters}

\noindent
where

\begin{mathletters}
\begin{eqnarray}
&&
M = \left ( {1\over{\tilde V}} + \Pi_M \right )
\Big [ \Pi_{\theta\theta}\tilde 
\Pi_{\varphi\varphi} - \Pi_{\theta\varphi} \Pi_{\varphi\theta} \Big ],
\label{detsym}
\\
&&
\nonumber\\
&&
\tilde \Pi_{\varphi\varphi} = \Pi_{\varphi\varphi} + {{\Pi_M}\over{1+\tilde V 
\Pi_M}} \quad , \quad
\tilde V = 4\pi e^2 d .
\end{eqnarray}
\end{mathletters}

\noindent
Equations \ (\ref{propsym}) are written for the case $Qd\ll 1$.

Substituting the polarization operators Eqs.\ (\ref{poop}) 
and \ (\ref{polm}) into Eqs.\ (\ref{propsym}), we obtain

\begin{mathletters}
\begin{eqnarray}
&&
{\cal D}^{\varphi\varphi}_M = - {{\nu_s^2}\over{M}}
\left [ {{\omega^2}\over{\tilde V \nu_s}} - \pi\Delta D_s Q^2
\left (1+{1\over{\tilde V \nu_s}} \right ) \right], \\
&&
\nonumber\\
&&
{\cal D}^{\varphi\theta}_{MS} = - {1\over{M}}
{{i\omega\nu_s}\over{\tilde V}}, \\
&&
\nonumber\\
&&
{\cal D}^{\theta\theta}_S = - {1\over{M}} \; {1\over{\tilde V}} \;
{{\zeta D_m Q^2 - i\omega\nu_s}\over{-i\omega + D_m Q^2}},
\end{eqnarray}
\label{propags}
\end{mathletters}

\noindent
where $\zeta = \nu_m + \nu_s ( 1 + \nu_m \tilde V)$ and 
Eq.~ (\ref{detsym}) takes the form

\begin{mathletters}
\begin{eqnarray}
&&
M = - \pi\Delta {{\nu_s}\over{\tilde V}} D_s \zeta
{{D_m Q^2}\over{-i\omega + D_m Q^2}} P(Q^2), \\
&&
\nonumber\\
&&
P(Q^2) = Q^2 - {{i\omega\nu_s}\over{\zeta D_m}} -
{{\omega^2\nu_m}\over{\pi\Delta\zeta D_s}}.
\label{pole}
\end{eqnarray}
\label{det}
\end{mathletters}

The pole $P(Q^2)$ in the fluctuation propagators means that there is
a well-defined collective mode with linear dispersion relation. 
This mode is the phason, discussed in Sec~\ref{col}. 
The imaginary term in Eq.~(\ref{pole}) determines the finite
lifetime of the phason. 
However, this term is small, since $D_m/D_s \gg 1$, 
and the phason lifetime is large by this parameter.

We now combine these propagators into a single 
retarded propagator of the field $\phi$ using the definition of $\phi$

\begin{eqnarray}
\langle \phi \phi \rangle = \langle K[\varphi]K[\varphi] \rangle
- 2 \langle K[\varphi]\theta \rangle + \langle\theta\theta \rangle.
\end{eqnarray}

Substituting the explicit form Eq.\ (\ref{Kfunc}) 
of the linear functional of the Coulomb
fluctuations in the normal metal and the propagators 
Eq.\ (\ref{propags}) we find

\begin{eqnarray}
{\cal D}_{\phi\phi}^R (\omega; Q)&&= 
{{\tilde V}\over{\nu_s \pi\Delta D_s\zeta}}
 \;
{1\over{P(Q^2)}}
\Big [{\zeta\over{\tilde V}} + 
{{\nu_s}\over{\tilde V}} {{i\omega}\over{-i\omega
 + D_m Q^2}}
\nonumber\\
&&
\nonumber\\
&&
+ {{\nu_s^2}\over{D_m}}(1+{1\over{\nu_s\tilde V}})
{{\pi\Delta D_s}\over{-i\omega
 + D_m Q^2}} \Big ].
\label{dqw}
\end{eqnarray}

Note, that although each of the propagators Eq.\ (\ref{propags}) 
contains the $1/Q^2$ factor,
the combined propagator Eq.\ (\ref{dqw}) does not. 
This is a manifestation of the gauge invariance:
the zero mode here corresponds to a uniform perturbation 
with the constant scalar potential,
which can be gauged away and can not affect the physics.

Finally, to obtain the time-dependent propagator, which 
is needed to calculate the average
Eq.\ (\ref{avd-eps}) we integrate Eq.\ (\ref{dqw}) over the momentum. 
The last two terms in
brackets do not contribute by their analytic properties, 
while the first term gives for the
retarded propagator

\begin{equation}
{\cal D}_{\phi\phi}^R (\omega) = 
{{i {\rm sgn}\omega}\over{2\Delta G_s}}.
\label{dwr}
\end{equation}

\noindent
The corresponding Keldysh propagator is

\begin{equation}
{\cal D}_{\phi\phi}^K (\omega) = {{i {\rm sgn}\omega}\over{\Delta G_s}} 
N(\omega),
\label{dwk}
\end{equation}

\noindent
where the dimensionless conductance of the superconductor $G_s=2\pi\nu_s D_s$.
The distribution function $N(\omega)$ can be obtained 
from Eq.~(\ref{fw}) using the particular
form of the metallic distribution function 
$f_M(\epsilon)$ corresponding to the case when
the system is driven out of equilibrium by the applied voltage $U$

\begin{equation}
f_M(\epsilon) = -{1\over{2}} 
\left [ \eta \left (\epsilon + {{eU}\over{2}} \right )
+ \eta \left (\epsilon - {{eU}\over{2}} \right ) \right ].
\label{distrU}
\end{equation}

\noindent
This results in the bosonic distribution function

\begin{eqnarray}
\omega N(\omega) \; = \; \left \{
\matrix{
|\omega |                 & \; , \; \; {\rm if} \; \; |\omega|>eU \cr
{1\over{2}}(eU+|\omega |) & \; , \; \; {\rm if} \; \; |\omega|<eU \cr
}
\right. .
\label{df}
\end{eqnarray}

\subsection{Results for tunneling conductance.}
\label{results}

Using the propagators Eqs.\ (\ref{dwr}) and \ (\ref{dwk}) 
we now calculate the average
Eq.\ (\ref{avd-eps}). 
First, let us separate the 
equilibrium ($V=0$) and non-equilibrium contributions

\begin{equation}
{\cal D}_{\phi\phi}^{+-}(t) - {\cal D}_{\phi\phi}^{+-}(0) = 
{\cal D}_{ne} + {\cal D}_{eq}.
\end{equation}

\noindent
The non-equlibrium part comes entirely from the 
Keldysh propagator Eq.\ (\ref{dwk}):

\begin{eqnarray}
{\cal D}_{ne}(t) &&  = {1\over{2}}
\Big [ {\cal D}_{\phi\phi}^K (t) - {\cal D}_{\phi\phi}^K (0) \Big ] _{eU} 
\nonumber\\
&&
\nonumber\\
&&
- {1\over{2}} \Big [ {\cal D}_{\phi\phi}^K (t) - {\cal D}_{\phi\phi}^K (0) 
\Big
] _{eU=0}.
\end{eqnarray}

\noindent
Taking the Fourier transform we get

\begin{equation}
{\cal D}_{ne}(t) = {{i eU}\over{2\pi\Delta G_s}} \left [ S_1(eUt)
+{{\sin eUt}\over{eUt}} - 1 \right ],
\label{dne}
\end{equation}

\noindent
where $S_1(z)=\ln z + \gamma - {\rm Ci}(z)$, ${\rm Ci}(z)$ 
is the integral cosine function \cite{abr}
and $\gamma = 0.577\dots$ is the Euler constant.

The remaining equilibrium part is determined by the Fourier transform of

\begin{equation}
{\cal D}_{eq}(t) = {1\over{2}}
\Big [ {\cal D}_{\phi\phi}^K (t) - {\cal D}_{\phi\phi}^K (0) \Big ] _{eU=0} +
{1\over{2}} {\cal D}_{\phi\phi}^R (t),
\label{deq}
\end{equation}

\noindent
which is given by the integral

\begin{eqnarray}
{\cal D}_{eq}(t) = && {i\over{\Delta G_s}} \int\limits_0^\infty
{{d\omega}\over{2\pi}} \left (1-e^{-i\omega t} \right )\nonumber\\
&&
\nonumber\\
&&
\approx {i\over{2\pi G_s}} - {1\over{G_s\Delta t}}; 
\quad t\gg {1\over{\Delta}}.
\label{deqi}
\end{eqnarray}

\noindent
Here we have used the fact that the advanced propagator 
vanishes at $t>0$.

It is easy to see that when substitued into Eq.\ (\ref{avd-eps}), 
only the non-equlibrium
propagator Eq.\ (\ref{dne}) gives a noticable 
deviation from the non-interacting (Golden-rule-type) behaviour. 
Consider Eq.~(\ref{avd-eps}) in the absence of interactions. 
Then it is the BSC Green's function given by Eq.\ (\ref{bcs}), 
which at frequencies near the gap
diverges as $\sim 1/\sqrt{|\Delta-\epsilon|}$. 
The equilibrium correction Eq.\ (\ref{deqi})
at long times behaves like $1/(\Delta t)$ 
(since the upper integration limit in Eq.\ (\ref{deqi}) is cut off by the gap). 
Expanding the exponential will only give rise to extra
factors of $|\Delta-\epsilon|$ in the numerator, thus being small. 
Therefore we can neglect the equilibrium correction.

The non-equilibrium contribution in Eq.\ (\ref{avd-eps}) 
is thus dominant and the average
Green's function Eq.\ (\ref{sfunc-av}) at $\epsilon >0$ is

\begin{eqnarray}
{\rm Re}\bar h_S^R = \sqrt{{\Delta\over{|\Delta-\epsilon|}}}
F\left({{eU}\over{\epsilon-\Delta}}, {{eU}\over{\pi\Delta G_s}}\right),
\label{ri}
\end{eqnarray}

\noindent
where $F$ is a function of two dimensionless variables 
$y=eU/(\epsilon - \Delta)$ and
$z=eU/\pi\Delta G_s$.
It is given by

\begin{eqnarray}
F(y;z)={\rm Re}\int\limits_0^\infty {{dx}\over{\sqrt{i\pi x}}}
e^{ix {\rm sgn}y} e^{-z[S_1(|y|x) + {{\sin yx}\over{yx}} -1]}.
\label{F}
\end{eqnarray}

The tunneling conductance Eq.\ (\ref{cond-av}) depends on 
the real part of the Green's function Eq.\ (\ref{ri}). 
The external voltage in the experimental setting is always
less than the superconducting gap, so that $z \ll 1$. 
To probe the vicinity of the gap
we need to look at $\epsilon \sim \Delta$ corresponding 
to the limit $y \gg 1$. In that limit Eq.\ (\ref{F}) gives

\begin{eqnarray}
F(y; z) = |y|^{-z} \left \{
\matrix{
{{\pi z}\over{2}}   & \; , \; \; {\rm if} \; \; y<0  \cr
1                   & \; , \; \; {\rm if} \; \; y>0
}
\right., \quad |y| \gg 1.
\label{r}
\end{eqnarray}

That describes the behaviour of the density of states 
in the superconductor in the vicinity
of the gap edge, 
in particular the appearance of non-zero density
of states inside the gap ($y <0 $ result)

For completeness we include the solution in the opposite 
limit $y \ll 1$ which describes the tail
of the density of states as we go deeper into the gap. 
When $\epsilon > \Delta$ we can expand the exponential to obtain

\begin{equation}
F(y; z) = 1 + {3\over{16}} z y^2,\quad y>0;\  y\ll 1.
\label{rtail}
\end{equation}

\noindent
However, deep in the gap the perturbation theory gives zero 
(due to energy conservation :
no single particle process is possible 
at $\epsilon < \Delta - eU$).
Therefore, we have to evaluate the full integral. 
This yields the non-perturbative result, which
describes the multi-phason processes

\begin{equation}
F(y; z) \approx
{1\over{\sqrt{2}}} \left({{|y|ze}\over{2}}\right)^{1\over{|y|}}
\left(\ln {2\over{|y|z}}\right)^{{4-|y|}\over{2|y|}}, \quad y<0,\ |y|\ll 1.
\label{rtail2}
\end{equation}

\noindent
Here $e=2.718\dots$.

Finally, we express the result for the tunneling conductance 
Eq.\ (\ref{cond-av}) in terms
of the function $F$

\begin{equation}
G_T = {1\over{2}} G_0 \left| 
{\Delta\over{\Delta-|\epsilon|}}\right|^{{1\over{2}
}}
\sum\limits_{\epsilon=eV\pm eU/2}
F\left({{eU}\over{|\epsilon|-\Delta}}, {{eU}\over{\pi\Delta G_s}}\right),
\label{cond-res}
\end{equation}

\noindent
where $F$ is given by Eq.~(\ref{F}) and its asymptotic behavior 
is described by Eqs.~(\ref{r})-(\ref{rtail2}).

\section{Conclusions}

In this paper we have studied the effect of electron-electron
interactions on the tunneling conductance of N-S junction with the
metallic layer driven out of equilibrium. 
Contrary to the common belief, the tunneling current 
can not be presented as
a convolution of bare BCS density of states and the derivative of the
non-equlibrium distribution function in metal. 
We have demonstrated, that the
interaction between the tunneling electrons and those in the normal 
metal drastically affect this current modifying
the superconducting density of states.  
This modification is manifested in particular 
in appearance of finite density of
states at energies even smaller than $\Delta$, see Sec.~\ref{results}.  
This effect can complicate the clear determination of the distribution
function by tunneling experiments similar to Ref.~\onlinecite{exp1}.

The reason for the strong effect of the non-equlibrium state in the
metal on the superconducting density of states is apperance of
low-energy collective mode on the N-S junction, analogous to
the Schmidt-Sch\"on mode or the second sound,
which we called ``phason'', Sec~\ref{col}.
The tunneling can be accompanied by
emission or absorbtion of phasons. The shot noise in the normal layer 
generates these phasons, thus, affecting the
tunneling density of states.

The particular form of the phason spectrum Eq.~(\ref{sp}) and 
consequently the result for the tunneling conductance
Eq.~(\ref{cond-res}) depend on the geometry of the junction. In this paper
we considered the simplest model of the junction, namely the 2D
sandwich. Calculations of the effect of the phason assisted tunneling 
on various experimental realizations of the N-S junction are outside
of the scope of this paper. Therefore our results can not be expected 
to describe the experiment \cite{exp1} in detail. However, we have 
demonstrated an existence of a microscopic mechanism changing
the  tunneling conductance, which is different from
the trivial broadening of the distribution function
in the normal metal. 
To test which mechanism dominates in reality 
we have suggested the independent experiment, see
Sec.~\ref{sec:expt}.

\acknowledgements
We are thankful to L.S. Levitov for discussions on the early stage 
of this work, which was begun in ICTP Trieste.
Helpful conversations with A.V. Andreev,
L.I. Glazman, A.I. Larkin, A.J. Leggett, A.J. Millis, and M.Yu. Reyzer  
are gratefully acknowledged.
I.A. is A.P. Sloan and Packard research fellow.
The work at Princeton University was supported by  by ARO Grant
DAAG55-98-1-0270.

\appendix

\section{}

Here we demonstrate the choice Eq.\ (\ref{Kfunc}) of the functional 
$K[\varphi]$ in  
Eq.\ (\ref{mfunc}). We simply substitute the Green's function 
Eq.\ (\ref{mfunc}) into the 
Usadel equation Eq.\ (\ref{usm}) and require that in equilibrium 
the correction 
${\bf g}_{M1}(t_1 , t_2)$ is porportional to the square of the gradient
${{\bf g}_{M1} \sim (\nabla K[\varphi])^2}$ and out of equilibrium to 
the variation of the 
distribution function 
${\bf g}_{M1} \sim \Delta K[\varphi] 
(f_M(\epsilon) - f_{M, eq}(\epsilon))$. 

To proceed with this program we start with the transformation

\begin{equation}
{\bf g}_M(t_1, t_2) = e^{iK[\varphi](t_1)\tau^3}  {\bf h}(t_1, t_2)
e^{-iK[\varphi](t_2)\tau^3}.
\label{tra}
\end{equation}

\noindent
After the substitution into the Usadel equation Eq.\ (\ref{usm})
we find the equation for the function ${\bf h}(t_1 , t_2)$

\begin{eqnarray}
- D_m\nabla^c_R &&( {\bf h} \circ \nabla^c_R {\bf h} )
+ {\bf \tau}^3 {\partial {\bf h} \over{\partial t_1}} +
 {\partial {\bf h} \over{\partial t_2}}{\bf \tau}^3 
\nonumber\\
&&
\nonumber\\
&& 
+i [{\partial K \over{\partial t}}- {\bf \Phi}, {\bf h} ]_t = 0,
\label{usmcor}
\end{eqnarray}

\noindent
where the covariant derivative now includes the commutator of 
the gradient of $K[\varphi]$
and the function ${\bf h}$

\begin{equation}
\nabla^c_R {\bf h} = \nabla_R {\bf h} + 
i \big [ \nabla_R K[\varphi] , {\bf h} \big ].
\end{equation}

\noindent
Here we assumed that the functional $K[\varphi]$ 
(which is a matrix in the Keldysh space) 
commutes with the scalar potential matrix ${\bf \Phi}$
which we will verify later using the explicit form of $K[\varphi]$.

We now treat Eq.\ (\ref{usmcor}) in perturbation theory 
${\bf h} = {\bf g}_{M0} + {\bf g}_{M1}$,
where ${\bf g}_{M0}$ is the free Green's function determined 
by Eq.\ (\ref{mfuncR0}). 
The equation becomes

\begin{eqnarray}
- D_m && \int dt_3 {\bf g}_{M0} (t_1-t_3) \Big [ 
\nabla^2 K[\varphi](t_3) \tau^3 {\bf g}_{M0} (t_3-t_2)\nonumber\\
&&
\nonumber\\
&& 
{ \; \; \; \; \; \; \; \; \; \; \; }  
- {\bf g}_{M0} (t_3-t_2) \tau^3 \nabla^2 K[\varphi](t_2) \Big ]
\nonumber\\
&&
\nonumber\\
&& 
+ \left( {{\partial K[\varphi]} \over{\partial t_1}}-{\bf \Phi}(t_1) \right )
 {\bf g}_{M0} (t_1-t_2) 
\nonumber\\
&&
\nonumber\\
&&
{ \; \; \; \; \; \; \; \; \; \; \; } 
- {\bf g}_{M0} (t_1-t_2) 
\left ( {{\partial K[\varphi]}\over{\partial t_2}} - {\bf \Phi}(t_2) \right )
\nonumber\\
&&
\nonumber\\
&& 
= {\cal F}\Big [ (\nabla K)^2; {\bf g}_{M1}; \nabla K \nabla 
{\bf g}_{M1} \Big ] .
\label{kfeq}
\end{eqnarray}

\noindent
The functional ${\cal F}$ in the right hand side of Eq.\ (\ref{kfeq}) 
(which explicit form is not needed for what follows) contains 
higher powers of $K[\varphi]$. We now choose such a form of
$K[\varphi]$ 
that all linear
terms (the left hand side of Eq.\ (\ref{kfeq})) cancel with the
external 
potential
${\bf \Phi}$. To do that we treat the left hand side of 
Eq.\ (\ref{kfeq}) as the equation
for $K[\varphi]$ (with zero in the right hand side).

To simplify the solution we notice that the non-interacting 
Green's function has the
matrix structure

\begin{equation}
{\bf g}_{M0} = \hat g_{M0} \otimes \tau^3,
\end{equation}

\noindent
where $\hat g_{M0}$ is the $2\times 2$ matrix in Keldysh space and 
$\tau^3$ acts in 
Nambu space. Therefore each term in the equation is proportional 
to $\tau^3$ which therefore
can be omitted. Then, using the normalization 
$ {\bf g}_{M0}\circ {\bf g}_{M0} = 1$ and 
performing the Fourier transform in space variable (only $K[\varphi]$ 
depends on $R$),
we obtain the equation for the functional $K[\varphi]$

\begin{eqnarray}
D_mQ^2 && \Big [ \int dt_3 \hat g_{M0} (t_1-t_3) K[\varphi](t_3) 
\hat g_{M0} (t_3-t_2) 
\nonumber\\
&&
\nonumber\\
&& 
{ \; \; \; \; \; \; \; } - K[\varphi](t_1) \delta (t_1 - t_2) \Big ]
\nonumber\\
&&
\nonumber\\
&& 
+ \left( {{\partial K[\varphi]} \over{\partial t_1}}- 
\hat \varphi(t_1) \right )
 \hat g_{M0} (t_1-t_2) 
\nonumber\\
&&
\nonumber\\
&&
{ \; \; \; \; \; \; \; } 
- \hat g_{M0} (t_1-t_2) 
\left ( {{\partial K[\varphi]}\over{\partial t_2}} - 
\hat \varphi(t_2) \right ) = 0,
\label{keq}
\end{eqnarray}

\noindent
where $\hat \varphi$ is the matrix

\begin{eqnarray}
\hat \varphi = 
\pmatrix
{
\varphi_+  & \varphi_-  \cr
\varphi_-  & \varphi_+  \cr
}_K ,
\label{fff}
\end{eqnarray}

\noindent
and the functional $K[\varphi]$ has the same structure

\begin{eqnarray}
K[\varphi] = 
\pmatrix
{
K_+  & K_-  \cr
K_-  & K_+  \cr
}_K .
\end{eqnarray}

\noindent
The Green's function $\hat g_{M0}$ is given by

\begin{eqnarray}
\hat g_{M0}(t_1 - t_2) =
\pmatrix
{
\delta (t_1 - t_2) & 2f(t_1 - t_2)      \cr
0                  & -\delta (t_1 - t_2) \cr
}_K,
\label{go}
\end{eqnarray}

\noindent
where the Keldysh component is related to the distribution function
(see Eq.\ (\ref{mfuncK0}) ). 

We substitute Eqs.\ (\ref{fff}) - \ (\ref{go}) into the matrix equation Eq.\ (\ref{keq}) 
and investigate each component separately.
The diagonal elements (``11'' and ``22'') of Eq.\ (\ref{keq}) differ only in sign 
and give the equation for $K_-$

\begin{equation}
D_mQ^2 K_- (t) - \left ( {{\partial K_-}\over{\partial t}} - \varphi_-(t) \right ) = 0,
\label{kmin}
\end{equation}

\noindent
which gives the solution in frequency domain (see  Eq.\ (\ref{Kfunc}) )

\begin{equation}
K_- = -{{\varphi_-}\over{i\omega + D_mQ^2}}.
\label{kms}
\end{equation}

This solution also satisfies the lower non-diagonal (``21'') element of Eq.\ (\ref{keq}).
The upper non-diagonal (Keldysh) element gives the equation for $K_+$ 
which we write in Fourier space

\begin{eqnarray}
&&{ \; \; \; \; \; } (D_m Q^2 -i\omega )K_+(\omega) [f(\epsilon) - f(\epsilon - \omega)]
\nonumber\\
&&
\nonumber\\
&&
+ 2 D_mQ^2  K_-(\omega)
\varphi_- [f(\epsilon)f(\epsilon - \omega) - 1]
\nonumber\\
&&
\nonumber\\
&&
{ \; \; \; \; \; \; \; } { \; \; \; \; \; \; \; } 
= \varphi_+ [f(\epsilon) - f(\epsilon - \omega)].
\label{kplu}
\end{eqnarray}

The last trick is to introduce the bosonic ``distribution function'' $N(\omega)$
Eq.\ (\ref{fw}). In the second term of Eq.\ (\ref{kplu}) we rewrite

\begin{eqnarray}
&&f(\epsilon)f(\epsilon - \omega) - 1 = - N(\omega)[f(\epsilon) - f(\epsilon - \omega)]
\nonumber\\
&&
\nonumber\\
&&
+ \Big [ N(\omega)[f(\epsilon) - f(\epsilon - \omega)] 
+ f(\epsilon)f(\epsilon - \omega) - 1 \Big ].
\label{nom}
\end{eqnarray}

In equilibrium the expression in square brackets is equal to zero, which is just a reflection
of the detailed balance principle. Then, using the solution Eq.\ (\ref{kms}) for $K_-$,
we obtain the solution

\begin{eqnarray}
K_+ = {{\varphi_+}\over{-i\omega + D_mQ^2}} 
- 2 \varphi_- {\rm Re}{{N(\omega)}\over{i\omega + D_m Q^2}},
\label{kps}
\end{eqnarray}

\noindent
which was previously given in the matrix form in Eq.\ (\ref{Kfunc}).

Out of equilibrium the solution Eq.\ (\ref{kps}) is not exact, since the detailed balance
principle is no longer valid and the square bracket in Eq.\ (\ref{nom}) gives rise to
corrections, which are proportional to the difference between equilibrium and non-equilibrium
distribution functions. These corrections should then be included in the next order
Green's function ${\bf g}_{M1}$. 

The calculation of the correction ${\bf g}_{M1}$ introduces the Altshuler-Aroonov corrections
to the conductity as well as the energy relaxation. We will not dwell on this issue in the
present paper.

\section{}

Here we show that the gauge transformation Eq.\ (\ref{sfunc}) captures the dominant effect
of phase and Coulomb fluctuations in the superconductor. Since we integrate the 
fluctuation propagators over momentum before using using them to calculate the tunneling
conductance, we can here set the momentum to zero from the very beginning taking the 
integrated Keldysh propagator Eq.\ (\ref{dwk}) as known. We only need the Keldysh propagator
since we are interested in non-equilibrium effects (see the paragraph following 
Eq.\ (\ref{deqi}). In the straightforward perturbation 
theory described in  Sec.~\ref{zm} the terms describing the phase fluctuations contained the
large factor of $\Delta$. These contributions can be taken into account
by means of the gauge transformation (similar to Eq.\ (\ref{sfunc}))

\begin{equation}
{\bf g}_S(t_1, t_2) = e^{i\theta_K(t_1)\tau^3} {\bf g} (t_1, t_2) e^{-i\theta_K(t_2)\tau^3},
\label{gt}
\end{equation}

\noindent
where now the function ${\bf g} (t_1, t_2)$ is not the BCS Green's function but is to be
determined from the Usadel equation, which after the transformation becomes

\begin{eqnarray}
-D_s\nabla^c({\bf g}\circ\nabla^c{\bf g}) 
+[{\bf H}_0, {\bf g}]_{\epsilon} + [{\bf \delta H}, {\bf g}]_{\epsilon} = 0,
\label{uegt}
\end{eqnarray}

\noindent
where

\begin{eqnarray}
[{\bf \delta H}, {\bf g}]_{\epsilon} = \int {{d\epsilon_3}\over{2\pi}} &&
\Big [{\bf \delta H} (\epsilon_1 - \epsilon_3) {\bf g} (\epsilon_3, \epsilon_2)\nonumber\\
&&
\nonumber\\
&&
 -
{\bf g} (\epsilon_1, \epsilon_3){\bf \delta H} (\epsilon_3 - \epsilon_2)\Big ] ,   
\end{eqnarray}

\noindent
the covariant derivative after the gauge transformation includes a commutator with the gradient
of $\theta$

\begin{equation}
\nabla^c{\bf g} = \nabla{\bf g} + i [ \nabla\theta_K\tau^3,{\bf g}],
\label{der}
\end{equation}

\noindent
and ${\bf \delta H} = {\bf \Phi} + \dot {\bf \Theta}$. 

In addition to the equation Eq.\ (\ref{uegt}) the function ${\bf g} (t_1, t_2)$ has to 
satisfy the constraint

\begin{equation}
{\bf g} \circ {\bf g} = \hat {\bf 1} \delta(t_1 - t_2),
\label{constr}
\end{equation}

\noindent 
which fixes the normalization.

Now instead of $\Delta$ the phase fluctuation term has a factor of $\omega$. In what follows
we show that all the corrections calculated by perturbation theory will be small at
least as a power of $\omega/\Delta$ and thus the gauge transformation indeed captures the
dominant contribution. Since we already have the factor of $\omega$ n the numerator, the only 
possibility to obtain a strong correction is to excite a soft mode. Therefore 
to simplify the discussion we first separate the gapped modes which can not give rise to
strong corrections (since we are interested in energies much smaller than $\Delta$) and then
treat the remaining soft modes in perturbation theory. 

The separation of the modes can already be seen in the absence of fluctuations.
The free (BCS) solution is given by Eq.\ (\ref{bcs}) and can be written in the following 
way

\begin{eqnarray}
g_{S0}^{R}(\epsilon) \; && = {{\epsilon\sigma_3 
+ \Delta\sigma_2}\over{\sqrt{(\epsilon \pm i0)^2 - \Delta^2}}}
\nonumber\\
&&
\nonumber\\
&&
\approx \sqrt{{\Delta\over{(\epsilon \pm i0) - \Delta}}} 
\Big [(\sigma_3 + \sigma_2) 
\nonumber\\
&&
\nonumber\\
&&
{ \; \; \; \; \; \; \; } { \; \; \; \; \; \; \; } { \; \; \; \; \; \; \; } 
+ {{\epsilon - \Delta}\over{\Delta}}(\sigma_3 - \sigma_2) \Big ],
\label{bcsn}
\end{eqnarray}

\noindent
where only in this section we denote by $\sigma_2$ and $\sigma_3$ 
the following matrices in the Nambu space

\begin{eqnarray}
\sigma_2 =
\pmatrix
{
 0  &  1  \cr
-1  &  0  \cr
}_N 
\quad ; \quad
\sigma_3 =
\pmatrix
{
1  &  0  \cr
0  & -1  \cr
}_N .
\end{eqnarray}

\noindent
Here the large and small parts of the solution (in the
limit ${\epsilon - \Delta}\ll\Delta$ we are interested in) turned out to have
different matrix structure. 

To see this structure in the equation we first explicitly separate the fast oscillating
part of the Green's functions writing the solution as

\begin{equation}
{\bf g} (t_1, t_2) = e^{i\Delta (t_1- t_2)} {\bf g}_1 (t_1, t_2).
\label{osc}
\end{equation}

\noindent
Here we are focusing on positive energies.
 
Now the functions ${\bf g}_1$ varies slowly in time (since the external potential 
${\bf \delta H}$ is a slow function) and we get the equation

\begin{equation}
-D_s\nabla^c({\bf g}_1\circ\nabla^c{\bf g}_1) +
[{\bf \tilde H}, {\bf g}_1]_{\epsilon} + [{\bf \delta H}, {\bf g}_1]_{\epsilon} = 0,
\label{feq}
\end{equation}

\noindent 
where the Hamiltonian ${\bf \tilde H}$ has an additional term

\begin{equation}
{\bf \tilde H} = {\bf H}_0 + i\Delta{\bf 1}_K\otimes\sigma_3.
\label{hami}
\end{equation}

We now see that the large term in the BCS Green's function Eq.\ (\ref{bcsn}) commutes 
with the Hamiltonian  Eq.\ (\ref{hami}) (neglecting the $\epsilon$ term for the moment),
while the small term anticommutes. In the same manner we will look for a solution of 
Eq.\ (\ref{feq}) which contains matrices commuting and anticommuting with ${\bf \tilde H}$

\begin{eqnarray}
{\bf g}_1 = \hat G_1 \otimes {\bf 1}_N +&& \hat G_2 \otimes (\sigma_3 + \sigma_2)
\nonumber\\
&&
\nonumber\\
&&
+ \hat H_1 \otimes \sigma_1 + \hat H_2 \otimes (\sigma_3 - \sigma_2),
\label{matr}
\end{eqnarray}

\noindent
where hat denotes matrices in Keldysh space, while the $\sigma_j$ act in Nambu space.
The anticommuting terms $\hat H_j$ are small and the commuting terms $\hat G_j$ are large. 
In the absence of fluctuations $\hat G_1=\hat H_1=0$ and $\hat G_2$ and $\hat H_2$ form 
the BCS Green's function Eq.\ (\ref{bcsn}). 

When we substitute the solution Eq.\ (\ref{matr}) into the transformed Usadel equation
Eq.\ (\ref{feq}), in the first order we only use the non-commuting terms $\hat H_j$
in the commutator with the Hamiltonian, neglicting the smaller contribution from the 
other terms in the equation. As a result we obtain

\begin{eqnarray}
(\epsilon_1&&-\epsilon_2)[ \hat G_1\otimes \sigma_3 + \hat G_2 \otimes {\bf 1}_N] +
(\epsilon_1+\epsilon_2) \hat G_2 \otimes \sigma_1 
\nonumber\\
&&
\nonumber\\
&&
+ [{\bf \delta H}, \hat G_1]_{\epsilon}\otimes {\bf 1}_N
+ [{\bf \delta H}, \hat G_2]_{\epsilon}\otimes (\sigma_3 + \sigma_2)
\nonumber\\
&&
\nonumber\\
&&
+ 2 \Delta \hat H_1 \otimes (\sigma_3 + \sigma_2)
- 4 \Delta \hat H_2 \otimes \sigma_1 + \cdots = 0,
\end{eqnarray}

\noindent
where $\cdots$ denote gradient terms which we do not include here due to their complexity but
will restore in the next equation.

Collecting terms with the matrix structure corresponding to four matrices in 
Eq.\ (\ref{matr}) we obtain the final equations for $\hat H_j$ and $\hat G_j$

\begin{mathletters}
\begin{eqnarray}
&&
(\epsilon_1-\epsilon_2)\hat G_1 - D_s \nabla A_4 
\nonumber\\
&&
\nonumber\\
&&
{\; \; \; \; \; \; \; \; \; \; \; \; \; } 
-  {i\over{2}}D_s [\nabla\theta_K, (A_1+A_2)] = 0,
\label{e1}
\\
&&
\nonumber\\
&&
(\epsilon_1-\epsilon_2)\hat G_2 + [{\bf \delta H}, \hat G_1]_{\epsilon}
- D_s  \nabla A_1 - iD_s  [\nabla\theta_K, (A_4+A_3)] 
\nonumber\\
&&
\nonumber\\
&&
{\; \; \; \; \; \; \; \; \; \; \; \; \; } 
+2D_s  [\nabla\theta_K, \hat G_2\circ\nabla\theta_K \hat G_2]  = 0,
\\
&&
\nonumber\\
&&
[{\bf \delta H}, \hat G_2]_{\epsilon} + 2 \Delta \hat H_1 - {i\over{2}}D_s  [\nabla\theta_K, A_1] 
- {i\over{2}}D_s  \{\nabla\theta_K, A_2\}
\nonumber\\
&&
\nonumber\\
&&
{\; \; \; \; \; \; \; \; \; \; \; \; \; } 
 -D_s \nabla (A_3+2i\hat G_2\circ\nabla\theta_K \hat G_2) = 0,
\\
&&
\nonumber\\
&&
(\epsilon_1+\epsilon_2)\hat G_2 - 4 \Delta \hat H_2 
-D_s  \nabla A_2 + iD_s  \{\nabla\theta_K, A_4\} 
\nonumber\\
&&
\nonumber\\
&&
{\; \; \; \; \; \; \; \; \; \; \; \; \; } 
-iD_s [\nabla\theta_K, (A_3+2i\hat G_2\circ\nabla\theta_K \hat G_2)]  = 0.
\end{eqnarray}
\label{foq}
\end{mathletters}

\noindent
Here the curly brackets denote anticommutators. 
The matrices $A_j$ are the gradient terms 
porportional to $\hat G_1$

\begin{mathletters}
\begin{eqnarray}
&&
A_1 = \hat G_1 \circ \nabla \hat G_1 + i \hat G_1 \circ 
[\nabla\theta_K, \hat G_2]
\nonumber\\
&&
\nonumber\\
&&
{\; \; \; \; \; \; \; \; \; \; \; \; \; } 
-i \hat G_2 \circ [\nabla\theta_K, \hat G_1],
\\
&&
\nonumber\\
&&
A_2 = - i \hat G_1 \circ [\nabla\theta_K, \hat G_2]
-i \hat G_2 \circ \{\nabla\theta_K, \hat G_1\},
\\
&&
\nonumber\\
&&
A_3 = \hat G_1 \circ \nabla \hat G_2 + \hat G_2 \circ 
\nabla \hat G_1 + A_4,
\\
&&
\nonumber\\
&&
A_4 = {i\over{2}}\hat G_1\circ [\nabla\theta_K, \hat G_1].
\end{eqnarray}
\label{aaa}
\end{mathletters}

Since all the terms in Eq.\ (\ref{e1}) are proportional to 
$\hat G_1$ it is still not
generated in the first order, therefore all the $A_j$ 
terms in the first order are equal to zero.
The only correction from the gradients comes from the term 
quadratic in $\hat G_2$,
which changes the equation for $\hat G_2$ in the presence of 
phase fluctuations

\begin{equation}
(\epsilon_1-\epsilon_2)\hat G_2 
-2 [\nabla\theta_K, \hat G_2\circ\nabla\theta_K \hat G_2]  = 0.
\label{g22}
\end{equation}

\noindent
However this is the equilibrium correction: for the only term coupling
to the non-equilibrium
distribution function, the classical field $\theta_+$, the commutator 
is equal to zero.
In any case, at zero frequency the correction is equal zero and thus 
it does not produce any
additional singularity. Thus 
the first order corrections are small by a factor of $1/\Delta$

\begin{mathletters}
\begin{eqnarray}
&&
\hat G_1 = 0,
\\
&&
\nonumber\\
&&
G_2^{R, A} = {1\over{2}}\sqrt{{\Delta\over{(\epsilon \pm i0)}}},
\\
&&
\nonumber\\
&&
H_2^{R, A} = {1\over{2}}{{\epsilon}\over{\Delta}}
\sqrt{{\Delta\over{(\epsilon \pm i0)}}} ,
\\
&&
\nonumber\\
&&
\hat H_1 = - {1\over {2\Delta}} [{\bf \delta H}, \hat G_2]_{\epsilon} 
+ 2i D_s \hat G_2\circ\nabla^2\theta_K \hat G_2,
\end{eqnarray}
\end{mathletters}

\noindent
where $\epsilon$ is counted from $\Delta$ after the transformation 
Eq.~(\ref{osc}).

Thus we see that ${\bf \delta H}$ does not couple soft modes to 
each other and so it can 
introduce only perturbative corrections of the order 
$({\bf \delta H}/\Delta)^2$.
Therefore in the leading order in $1/\Delta$ the dominant effect of the phase
fluctuations is indeed captured by the gauge transformation Eq.~(\ref{sfunc}).

\section{}

Here we perform the averaging over the fluctuating field in 
Eq.\ (\ref{sfunc-av}).
The BCS Green's function $g_{S0}$ has the following structure 
in frequency domain

\begin{mathletters}
\begin{eqnarray}
&&g_{S0}^{R(A)}(\epsilon) = { {\hat H_0}\over{
\sqrt{(\epsilon \pm i0)^2 - \Delta^2}}},\\
&&
\nonumber\\
&&
g_{S0}^K (\epsilon) = \tanh{\epsilon\over{2T}} ( g_{S0}^R - g_{S0}^A ),
\end{eqnarray}
\label{bcs}
\end{mathletters}

\noindent
where the matrix in Nambu space is

\begin{eqnarray}
\hat H_0 =  
\pmatrix
{
\epsilon  & \Delta    \cr
-\Delta   & -\epsilon \cr
}_{SC}.
\end{eqnarray}

\noindent
In time domain the matrix structure is the same and we can represent 
$g_{S0}$ as a sum
of two parts

\begin{equation}
g_{S0} = h_0 \tau^3 + l_0 \tau^2.
\end{equation}

\noindent
To calculate the tunneling current we only need the normal component 
of $g_{S0}$, therefore
we can focus on averaging $h_0$. Then the matrix $\tau^3$ can be 
carried through, so that

\begin{equation}
\bar h_0 = \langle e^{i\phi(t_1)} h_{0}(t_1-t_2) e^{-i\phi(t_2)} \rangle.
\end{equation}

Since only the retarded function enters the expression for the 
tunneling current 
Eq.\ (\ref{cond-av}), we now multiply the Keldysh space matrices to 
get $\bar h_0^R$.
The Green's function has it's usual (rotated) form

\begin{eqnarray}
h_0 = 
\pmatrix
{
h_0^R  & h_0^K  \cr
0      & h_0^A  \cr
}_K ,
\end{eqnarray}

\noindent
while the fluctuating field is

\begin{eqnarray}
\phi = 
\pmatrix
{
\phi_+  & \phi_-  \cr
\phi_-  & \phi_+  \cr
}_K .
\end{eqnarray}

\noindent
After matrix multiplication we get for the retarded function

\begin{eqnarray}
\bar h_0^R(t_1-t_2) = && \langle e^{i(\phi_+(t_1)-\phi_+(t_2))} \Big [
h_0^R \cos\phi_-(t_1)\cos\phi_-(t_2) \nonumber\\ 
&&
\nonumber\\
&&  
- h_0^K \cos\phi_-(t_1) i\sin\phi_-(t_2) \nonumber\\
&&
\nonumber\\
&&  
+h_0^A \sin\phi_-(t_1)\sin\phi_-(t_2) \Big ]
\label{av3}
\end{eqnarray}

By definition a retarded Green's function is only non-zero when its 
time argument is
positive. In that region an advanced Green's function is
zero. Therefore 
we can 
simplify Eq.\ (\ref{av3}) as

\begin{eqnarray}
\bar h_0^R&&(t_1>t_2) = {1\over{2}}
\langle e^{i(\phi_+(t_1)-\phi_+(t_2))} \cos\phi_-(t_1)\nonumber\\ 
&&
\nonumber\\
&&  
\Big [ (h_0^R + h_0^K) e^{-i\phi_-(t_2)} + (h_0^R - h_0^K) 
e^{i\phi_-(t_2)} \Big ].
\label{av4}
\end{eqnarray}

We are now ready to perform the average:

\begin{mathletters}
\begin{eqnarray}
\langle && e^{i\big (\phi_+(t_1)-\phi_+(t_2)\big )
+i\big (\phi_-(t_1)-\phi_-(t_2)\big )} \rangle =
{\; \; \; \; \; \; \; \; \; \; \; \; \; \; \; \; }
\nonumber\\ 
&&
\nonumber\\
&&  
{\; \; \; \; \; \; \; \; \; \; \; \; \; \; \; \; }
= e^{\langle \phi_+ \phi_+ \rangle  - \langle \phi_+^2 \rangle (0) +
\langle \phi_- \phi_+ \rangle  + \langle \phi_+ \phi_- \rangle},
\\ 
&&
\nonumber\\
\langle && e^{i\big (\phi_+(t_1)-\phi_+(t_2)\big )-
i\big (\phi_-(t_1)+\phi_-(t_2)\big )} \rangle =
{\; \; \; \; \; \; \; \; \; \; \; \; \; \; \; \; }
\nonumber\\ 
&&
\nonumber\\
&&  
{\; \; \; \; \; \; \; \; \; \; \; \; \; \; \; \; }
= e^{\langle \phi_+ \phi_+ \rangle - \langle \phi_+^2 \rangle (0) -
\langle \phi_- \phi_+ \rangle + \langle \phi_+ \phi_- \rangle},
\end{eqnarray}
\end{mathletters}

\noindent
where all the correlators $\langle \phi \phi \rangle$ depend on 
the time difference 
$t_1-t_2$, except where indicated.

We now introduce propagators ${\cal D}$ of the fluctuating fields

\begin{mathletters}
\begin{eqnarray} 
&&
\langle \phi_+ \phi_+ \rangle = i {\cal D}_{\phi\phi}^K (t_1-t_2), \\
&&
\nonumber\\
&&
\langle \phi_+ \phi_- \rangle = i {\cal D}_{\phi\phi}^R (t_1-t_2), \\
&&
\nonumber\\
&&
\langle \phi_- \phi_+ \rangle = i {\cal D}_{\phi\phi}^A (t_1-t_2) .
\end{eqnarray}
\end{mathletters}

\noindent
At $t_1>t_2$ the advanced propagator is zero. Therefore only sums and 
differences of the 
retarded and the Keldysh functions are present in Eq.\ (\ref{av4}). 
These can be expressed
in terms of functions of the original Keldysh basis as 
$h_0^K \pm h_0^R = 2h_0^{+- (-+)}$ 
and similarly with the fluctuation propagators to obtain 
the final expression in time
domain

\begin{eqnarray}
\bar h_S^R(t_1>t_2)&& = h_0^{+-}(t_1-t_2) 
e^{2i \big ({\cal D}_{\phi\phi}^{+-}(t_1-t_2) 
- {\cal D}_{\phi\phi}^{+-}(0)\big )} \nonumber\\
&&
\nonumber\\
&&
-  h_0^{-+}(t_1-t_2) 
e^{2i \big ( {\cal D}_{\phi\phi}^{-+}(t_1-t_2) 
- {\cal D}_{\phi\phi}^{-+}(0)\big )}.
\label{avd-time}
\end{eqnarray}

When Fourier transforming to the frequency domain, we notice that 
the first term in 
Eq.\ (\ref{avd-time}) contributes at positive frequencies, while 
the second term contributes
only at negative frequencies. Therefore

\begin{equation}
\bar h_S^R(\epsilon >0) = 
\int\limits_0^\infty dt e^{i\epsilon t} h_0^{+-}(t) 
e^{2i \big ( {\cal D}_{\phi\phi}^{+-}(t) 
- {\cal D}_{\phi\phi}^{+-}(0) \big )} .
\end{equation}

For reference purposes we list here the explicit form of the BCS 
function $h_0^{+-}$ in 
the time domain

\begin{equation}
h_0^{+-} (t>0) = {\Delta\over{\pi}} {\it K}_1(i\Delta t),
\end{equation}

\noindent
where ${\it K}_1$ is the modified Bessel function.

\end{multicols}

\end{document}